\documentclass[a4paper,11pt]{article}
\pdfoutput=1 

\usepackage{jcappub} 
\usepackage{aas_macros}

\usepackage[T1]{fontenc} 
\usepackage[table]{xcolor}
\usepackage{scalerel}
\usepackage{tikz}
\usetikzlibrary{svg.path}

\definecolor{orcidlogocol}{HTML}{A6CE39}
\tikzset{
  orcidlogo/.pic={
    \fill[orcidlogocol] svg{M256,128c0,70.7-57.3,128-128,128C57.3,256,0,198.7,0,128C0,57.3,57.3,0,128,0C198.7,0,256,57.3,256,128z};
    \fill[white] svg{M86.3,186.2H70.9V79.1h15.4v48.4V186.2z}
                 svg{M108.9,79.1h41.6c39.6,0,57,28.3,57,53.6c0,27.5-21.5,53.6-56.8,53.6h-41.8V79.1z M124.3,172.4h24.5c34.9,0,42.9-26.5,42.9-39.7c0-21.5-13.7-39.7-43.7-39.7h-23.7V172.4z}
                 svg{M88.7,56.8c0,5.5-4.5,10.1-10.1,10.1c-5.6,0-10.1-4.6-10.1-10.1c0-5.6,4.5-10.1,10.1-10.1C84.2,46.7,88.7,51.3,88.7,56.8z};
  }
}

\newcommand\orcid[1]{\href{https://orcid.org/#1}{\mbox{\scalerel*{
\begin{tikzpicture}[yscale=-1,transform shape]
\pic{orcidlogo};
\end{tikzpicture}
}{|}}}}

\title{On the consistency of \tpdf{\lcdm} with CMB measurements in light of the latest Planck, ACT and SPT data}

\author[a,1]{Rodrigo Calderón\orcid{0000-0002-8215-7292},\note{Corresponding author.}}
\author[a,b]{Arman~Shafieloo\orcid{0000-0001-6815-0337},}
\author[c,d,e]{Dhiraj Kumar Hazra\orcid{0000-0001-7041-4143}}
\author[a]{and Wuhyun Sohn\orcid{0000-0002-6039-8247}}

\affiliation[a]{Korea Astronomy and Space Science Institute,
Daejeon 34055, South Korea}
\affiliation[b]{University of Science and Technology,
Daejeon 34113, South Korea}
\affiliation[c]{The Institute of Mathematical Sciences, HBNI, CIT Campus, Chennai 600113, India}
\affiliation[d]{Homi Bhabha National Institute, Training School Complex, Anushakti Nagar, Mumbai 400085, India}
\affiliation[e]{INAF/OAS Bologna, Osservatorio di Astrofisica e Scienza dello Spazio,
Area della ricerca CNR-INAF, via Gobetti 101, I-40129 Bologna, Italy
}

\emailAdd{calderon@kasi.re.kr}
\emailAdd{shafieloo@kasi.re.kr}
\emailAdd{dhiraj@imsc.res.in}
\emailAdd{wuhyun@kasi.re.kr}

\abstract{
Using Gaussian Processes we perform a thorough, non-parametric consistency test of the \lcdm\ model when confronted with state-of-the-art TT, TE, and EE measurements of the anisotropies in the Cosmic Microwave Background by the \planck, ACT, and SPT collaborations. Using \lcdm's best-fit predictions to the TTTEEE data from \planck, we find no statistically significant deviations when looking for signatures in the residuals across the different datasets.  
The results of SPT are in good agreement with the \lcdm\ best-fit predictions to the \planck\ data, while the results of ACT are only marginally consistent. However, when using the best-fit predictions to CamSpec\textemdash a recent reanalysis of the \planck\ data\textemdash as the mean function, we find larger discrepancies between the datasets. Our analysis also reveals an interesting feature in the polarisation (EE) measurements from the CamSpec analysis, which could be explained by a slight underestimation of the covariance matrix.
Interestingly, the disagreement between CamSpec and \planck/ACT is mainly visible in the residuals of the TT spectrum, the latter favoring a scale-invariant tilt $n_s\simeq1$, which is consistent with previous findings from parametric analyses. We also report some features in the EE measurements captured both by ACT and SPT which are independent of the chosen mean function and could be hinting towards a common physical origin. For completeness, we repeat our analysis using the best-fit spectra to ACT+WMAP as the mean function. Finally, we test the internal consistency of the \planck\ data alone by studying the high and low-$\ell$ ranges separately, finding no discrepancy between small and large angular scales.
}

\begin{document}
\newcommand{\Cell}{\ensuremath{\mathcal{C}_\ell}}
\newcommand{\Dell}{\ensuremath{\mathcal{D}_\ell}}
\newcommand{\actpwmap}{\texttt{ACT+WMAP}}
\newcommand{\planck}{\emph{Planck}}
\newcommand{\chisq}{\ensuremath{\chi^2}}
\newcommand{\dlum}{\ensuremath{d_\text{L}}}
\newcommand{\dcom}{\ensuremath{d_\text{M}}}
\newcommand{\dang}{\ensuremath{d_\text{A}}}
\newcommand{\hMpc}[1]{\ensuremath{#1\,h^{-1}\mathrm{Mpc}}}
\newcommand{\lcdm}{$\Lambda$CDM}
\newcommand{\wcdm}{$w$CDM}
\newcommand{\Lkl}{\ensuremath{\mathcal{L}}}
\newcommand{\Om}{\ensuremath{Om}}
\newcommand{\Omm}{\ensuremath{\Omega_\text{m}}}
\newcommand{\Omo}{\ensuremath{\Omega_{\text{m},0}}}
\newcommand{\Omb}{\ensuremath{\Omega_\text{b}}}
\newcommand{\Omde}{\ensuremath{\Omega_\text{de}}}
\newcommand{\Oml}{\ensuremath{\Omega_\Lambda}}
\newcommand{\Omk}{\ensuremath{\Omega_k}}
\newcommand{\Omr}{\ensuremath{\Omega_\text{r}}}
\newcommand{\one}{\ensuremath{\mathbbm{1}}}
\newcommand{\Hord}{\ensuremath{H_0\rd}}
\newcommand{\vect}[1]{\ensuremath{\boldsymbol{#1}}}
\newcommand{\tens}[1]{\ensuremath{\mathbfss{#1}}}
\newcommand{\rd}{\ensuremath{r_\text{d}}} 
\newcommand{\diff} {\ensuremath{\mathrm{d}}} 
\newcommand{\diffn}[2] {\ensuremath{\diff^{#1}\vect{#2}}} 
\newcommand{\deriv}[2]{\ensuremath{\frac{\diff {#1}}{\diff {#2}}}}
\newcommand{\dpart}[2]{\ensuremath{\frac{\partial {#1}}{\partial {#2}}}}
\newcommand{\mean}[1]{\ensuremath{\left\langle #1 \right\rangle}}
\newcommand{\abs}[1]{\ensuremath{\left\lvert#1\right\rvert}}
\newcommand{\DD}{\ensuremath{\mathcal{D}}}
\newcommand{\expo}[1]{\ensuremath{\mathrm{e}^{#1}}}
\newcommand{\chsq}{\ensuremath{\chi^2}}
\newcommand{\seighto}{\ensuremath {\sigma_{8,0}}}
\newcommand{\hyperpars}{\ensuremath{(\sigma_f,\ell_f)}}


\def\tpdf#1{\texorpdfstring{#1}{Lg}}


\definecolor{lgray}{gray}{0.93}
\definecolor{deepmagenta}{rgb}{0.8, 0.0, 0.8}
\definecolor{ballblue}{rgb}{0.13, 0.67, 0.8}
\definecolor{RedWine}{rgb}{0.743,0,0}
\def\rcb#1{\textcolor{RedWine}{[RC:  #1]}}
\def\whs#1{\textcolor{teal}{[WS:  #1]}}
\maketitle
\flushbottom


\section{Introduction}

Observations stemming from the anisotropies in the Cosmic Microwave Background (CMB) have played a major role in establishing the standard model of Cosmology --- the \lcdm\ paradigm with power law primordial spectrum. Despite suffering from certain theoretical issues, this paradigm has been extremely successful in accounting for a wide variety of observations across many scales and epochs in the cosmic history. The \planck\ satellite provided the most precise estimation of its 6 main cosmological parameters to date \citep{Planck_results_2020,P18Cosmo2020}. In addition, other ground-based CMB experiments from the Atacama Cosmology Telescope (ACT) \citep{Aiola_2020,Choi_2020} and South Pole Telescope (SPT) \citep{Dutcher_2021,SPT-3G:2022hvq} collaborations have recently provided complementary measurements of temperature and polarisation of the CMB anisotropies. These focus on smaller, sub-degree angular scales (larger multipoles $\ell\gtrsim650$) and offer a new way of testing the robustness of \lcdm\ with higher-resolution CMB maps and independently of \planck.\\

Despite the success of the standard model, increasingly precise (low-redshift) measurements have reported a few statistically significant discrepancies \citep{Verde_2019,Cosmo_Intertwined_III_2021,Cosmo_Intertwined_II_2021}. The most notable example is the $\gtrsim5\sigma$ discrepancy in the value of the Hubble constant $H_0$, as measured by low-$z$ probes using the distance ladder \cite{Riess:2021jrx} and high-$z$ estimations, assuming \lcdm. 
A milder ($\sim2\sigma$) but longstanding discrepancy has also been reported between high and low-redshift estimations of the amplitude of matter fluctuations---characterized by $S_{8}\equiv\seighto\sqrt{\Omo/0.3}$---the latter preferring lower values compared to the early universe predictions \citep{HSC:2018mrq,2020A&A...634A.127A,2021Kids,DES:2021wwk,Amon_NL_S8,Ivanov_2023,survey2023des}.
The \lcdm\ model is also facing other (less-relevant) observational challenges, see \emph{e.g.} \citep{Bull:2015stt,Bullock:2017xww,Perivolaropoulos:2021jda,Bernal:2021yli}.\\

Moreover, CMB measurements are known to have mild inconsistencies between the different angular scales (high vs low multipoles \cite{Planck:2013pxb,Addison_2016}), and as measured by the different collaborations \citep{PhysRevD.103.063529,Di_Valentino_2022}. Indeed, even within \planck\ data alone, the TT spectrum seems to favor a lensing amplitude $A_{\rm L}>1$ \citep{P18Cosmo2020,Motloch:2018pjy} and provides ``evidence'' for a non-vanishing (positive) spatial curvature \citep{Planck:2015fie,P18Cosmo2020,DiValentino2019}, although it has been argued that these are purely stemming from statistical fluctuations \cite{Efstathiou_2021}; see also \citep{Handley_2021,Di_Valentino_2021,Vagnozzi_2021,https://doi.org/10.48550/arxiv.2210.09865} and references therein for further discussions on this. Furthermore, the results from the ACT collaboration seem to prefer a scale-invariant spectrum of primordial fluctuations, with $n_s\simeq1$, while \planck\ data excludes such a value at more than $3\sigma$ \citep{Ye_2021,Corona:2021qxl,Giare:2022rvg,PhysRevD.103.063529,https://doi.org/10.48550/arxiv.2210.06125,Jiang_2022}; see e.g. \citep{Schwarz:2015cma,Yeung:2022smn,Fosalba:2020gls} for other CMB anomalies. If these inconsistencies are not coming from systematics, they may hint towards new physics beyond the standard model.\\

In recent years, a lot of effort has gone into investigating extensions of \lcdm\ to try and provide physical explanations for some of the aforementioned discrepancies; see \emph{e.g.} \citep{Di_Valentino_2021_ItR,Shah_2021,Schoneberg:2021qvd,Abdalla:2022yfr} for a review. These often change the Universe's growth and/or expansion history at late times \citep{Pogosian2022,Heisenberg:2022gqk,Calderon:2020hoc}, or introduce new physics at early times such that (i) the physical size of the sound horizon $r_d\equiv r_s(z_d)$ decreases with respect to \lcdm\ \citep{Poulin:2018cxd,Niedermann:2019olb,ACT_EDE,NEDE_22,kamionkowski2022hubble}, (ii) the redshift of recombination is shifted  \citep{Jedamzik_2020,Galli_2022,https://doi.org/10.48550/arxiv.2107.02243,Sekiguchi_2021} or (iii) new features are introduced in the primordial spectrum of fluctuations \citep{Hazra:2022rdl,Antony:2022ert}. While appealing from the theoretical standpoint, very few of the proposed solutions are actually able to simultaneously address these tensions. For example, it has been argued that no late-time modification to $H(z)$ is able to raise the value of $H_0$ \citep{Knox:2019rjx,Keeley:2022ojz}, while modifications to the early universe  might create or exacerbate the tensions with low-$z$ observations; see \emph{e.g.} \citep{Ivanov:2020ril,Hill_2020,Smith:2020rxx,Murgia:2020ryi,Niedermann:2020qbw,DAmico:2020ods,2021CmPhy...4..123J,Simon:2022adh} for discussions on this topic. \\

Given the fundamental role of the CMB in cosmological analyses, it is crucial to understand whether the differences between the latest observations are coming from either statistical fluctuations, unaccounted systematics, or new physics beyond \lcdm. In this work, we test their statistical consistency using Gaussian Processes (GP) -- a non-parametric method that can effectively represent smooth deformations away from the model under consideration. If the differences between the datasets are entirely consistent with random fluctuations, then the GP regression should yield confidence levels consistent with zero. If not, then the GP can provide insights into what shape of deformation, either from systematics or theoretical inconsistency, is preferred by the data. \\

The structure of this paper is as follows. In Section \ref{MethodData}, we describe in detail the method and the data used in the analysis. We then proceed to perform the consistency tests using the best fit \lcdm\ predictions, by looking for structures in the residuals with respect to the mean function.  We start by confronting \lcdm\ to space-based observations by the Planck satellite. We use both the \texttt{Plik} and CamSpec likelihoods, which correspond to the \planck's official likelihood and the most recent reanalysis of the \planck\ data with the highest sky fraction, respectively, in Section \ref{LCDM_Planck}.
We then repeat the analysis using ground-based measurements by the Atacama Cosmology Telescope (ACT) and the South Pole Telescope (SPT) to look for discrepancies between the experiments in Section \ref{sec:ACTSPT}.  Finally, we test the robustness of our conclusions by using a different mean function in the analysis\footnote{Namely, we use \lcdm's best fit to the Planck 2018 official \texttt{plik} data release and \actpwmap\ data as alternative mean functions.} and investigate the consistency of the \planck\ data alone by studying the low-$\ell$ ($\ell<650$) and high-$\ell$ ranges ($\ell>650$) separately in Appendix \ref{sec:P18mean}, \ref{sec:ACTWMAPmean} and \ref{sec:HvsL}. In Appendix \ref{Calibration}, we assess whether an absolute scaling of the spectra (difference in calibration) can account for the mild inconsistencies between the different collaborations. The main findings of this analysis are summarized in \ref{sec:summary}. We then conclude and discuss future prospects in Section \ref{Conclusion}.

\section{Method and Data}\label{MethodData}

\subsection{Data}\label{sec:data}

\begin{figure}
    \centering
    \includegraphics[width=\columnwidth]{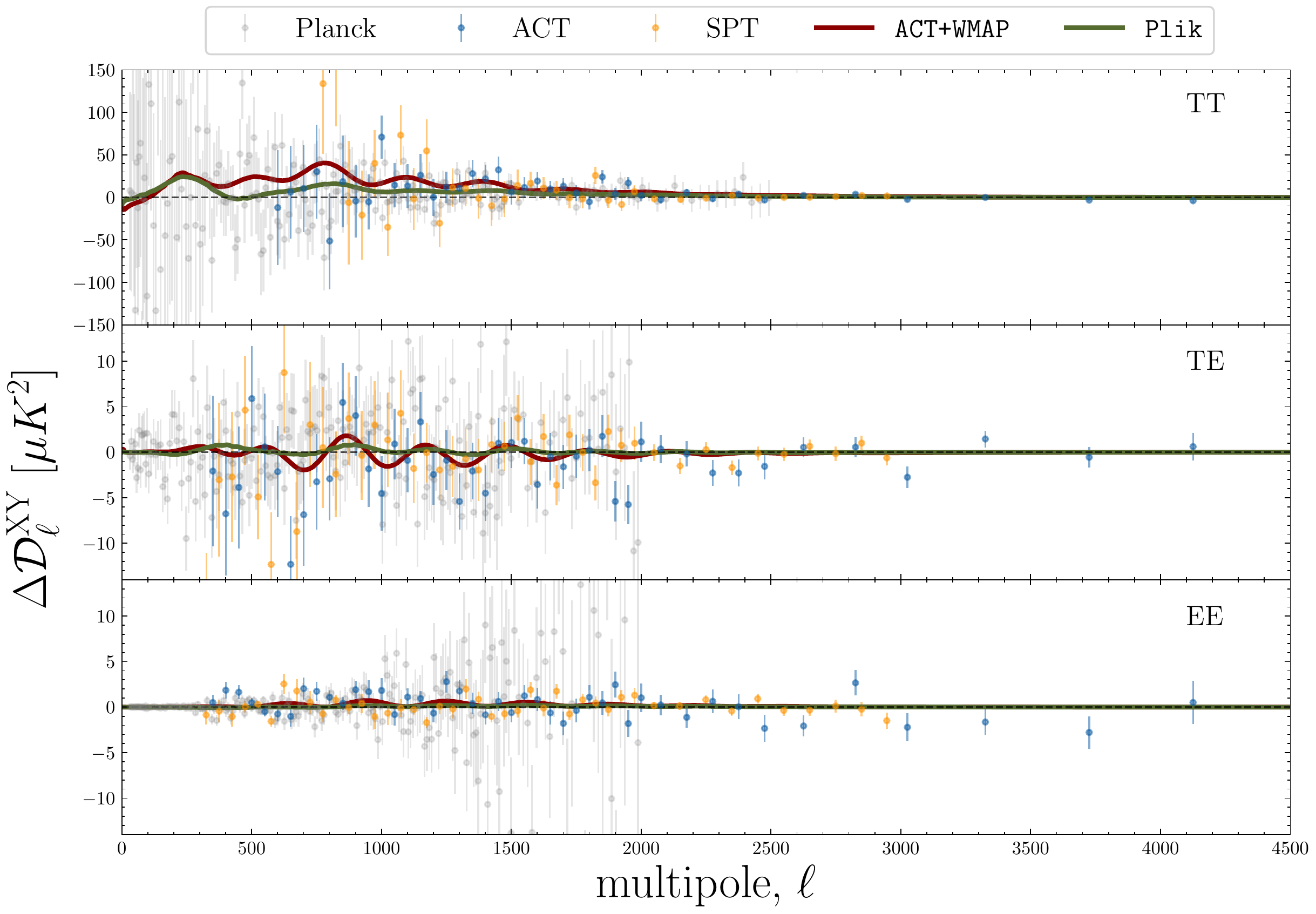}
    \caption{TT, TE and EE residuals with respect to \lcdm\ best fit spectra to CamSpec data. Solid red and green lines correspond to the residuals in the best fit predictions from \actpwmap\ and \texttt{Plik} with respect to the best fit to  CamSpec data, respectively.}
    \label{fig:residuals}
\end{figure}

In this work, we confront the predictions of \lcdm\ with different space and ground-based CMB experiments as a way of testing the consistency of the model and the robustness of the measurements. Namely, we consider the following data
\begin{itemize}
    \item \textbf{\planck\ 2018} - Temperature (TT), Polarisation (EE) and their cross-correlation (TE) from the final \planck\ 2018 data release  \citep{Planck_results_2020,P18Cosmo2020}. More specifically, we use the data from the \texttt{base-plikHM-TTTEEE-lowl-lowE-lensing} likelihood, which is publicly available at the Planck legacy archive\footnote{\url{https://pla.esac.esa.int/\#cosmology}}. These cover the range\footnote{We omit the $\ell<30$ measurements due to their non-Gaussian nature.} $\ell\in[30,2500]$ in TT and $\ell\in[30,2000]$ in TE and EE. We refer to this data simply as Planck 2018 (or \texttt{plik}). 
    \item \textbf{CamSpec PR4} - We also consider TT, TE and EE measurements from the latest \texttt{CamSpec NPIPE PR4\_v12.6} likelihood \cite{Rosenberg:2022sdy}; see \cite{Efstathiou_2021} for more details on \texttt{CamSpec}. These cover the range $\ell\in[30,2500]$ in TT and $\ell\in[30,2000]$ in TE and EE.  We refer to this data simply as CamSpec.
    \item \textbf{ACT DR4} - We use the  Atacama Cosmology Telescope (ACT) Temperature (TT), $E$-mode Polarisation (EE) and their cross-correlation (TE) from the \texttt{ACTPolliteDR4} likelihood in the latest ACT data release \citep{Aiola_2020,Choi_2020}. We refer to these measurements simply as ACT.
    \item \textbf{SPT 3G} - Finally, we also include the latest results from the South Pole Telescope Collaboration (SPT-3G) \cite{SPT-3G:2022hvq}. These are updated measurements of both E-mode Polarisation (EE) and Temperature-Polarisation cross-correlation (TE) from \cite{Dutcher_2021} but with the inclusion of TT measurements. These cover angular scales $\ell\in[750,3000]$ for TT and $\ell\in[350,3000]$ in TE and EE.
\end{itemize}

We should note that we use the minimum-variance-combined band powers for ACT and SPT data, which might not accurately reflect the full information contained in these datasets. We believe however that this can be seen as a zeroth-order approximation. A more rigorous analysis using the full (multi-frequency) likelihood might be needed for a more robust interpretation of the results.

\begin{table}[ht]
    \centering
    \caption{Differences in the \lcdm\ best-fit parameters for the different datasets.}
    \label{tab:mean_params}
    {\rowcolors{2}{lgray}{white}
    \begin{tabular}{ccccccc}
        \hline
		Best-fit & $H_0$ & $100\,\omega_b$ & $100\,\omega_c$ & $n_s$ & $\ln{(10^{10}A_s)}$ & $\tau$ \\ 
		\hline
		Planck 2018  & $67.32$  & $2.238$ & $12.01$& $0.9660$& $3.044$& $0.054$\\ 
		CamSpec  & $67.18$ & $2.218$ & $11.96$ & $0.9624$ & $3.035$ & $0.052$\\ 
		 $\actpwmap$  & $67.10$ & $2.223$ & $12.12$ & $0.9714$ &  $3.068$ & $0.061$\\ 

		\hline
    \end{tabular}}
\end{table} 

\subsection{Gaussian Process Regression}\label{Method}

Gaussian Processes (GP) \cite{rasmussen2006gaussian} have been extensively used in the literature to fit a smooth curve from noisy and/or sparse data without the need to write down an explicit parametric model \citep[see e.g.][]{PhysRevD.85.123530,PhysRevD.87.023520,Seikel_2012,Krishak:2021fxp,Calderon:2022cfj,Calderon:2023msm}. GP excels when the noise in the data is well approximated by a (multivariate) Gaussian distribution. It provides a posterior distribution of smooth functions given the data based on two assumptions on the functional form: the mean function ($\mu(x)$) and the kernel ($k(x,x')$). In-depth analyses of GP's dependence on these assumptions are given in \citep[e.g.][]{PhysRevD.85.123530,PhysRevD.87.023520,Hwang:2022hla}. The mean of the GP posterior distribution evaluated at a set of `test' points $\vect{x}_\star$ can be easily calculated through
\begin{equation}\label{mean_gp}
    \vect{\mu}= \vect{m}_\star+ \mathbf{K_\star}\mathbf{K}^{-1}(\vect{y}-\vect{m})
\end{equation}
where $\vect{m}_\star\equiv\mu(\vect{x}_\star)$, $\vect{m}\equiv \mu(\vect{x})$, $\mathbf{K}_\star \equiv k(\vect{x}_\star, \vect{x})$, $\mathbf{K}\equiv k(\vect{x},\vect{x})+\Sigma$, where the observations $\vect{y}$ are made at data points $\vect{x}$ with the data covariance matrix $\Sigma$. 
Similarly, the posterior of the covariance is obtained using
\begin{equation}\label{cov_gp}
    \vect{C}=\mathbf{K}_{\star\star}-\mathbf{K}_\star\mathbf{K}^{-1}\mathbf{K}^{T}_{\star}
\end{equation}

In practice, the calculation of such quantities amounts to a matrix inversion of $\mathbf{K}$. Computationally, a Cholesky decomposition is often preferred as it is a faster and numerically more stable procedure.
Finally, the log-marginal likelihood (LML) under a GP is given by
\begin{equation}\label{LML}
    \ln\mathcal{L}=-\frac12\big[\vect{r}^{T}\,\mathbf{K}^
    {-1}\,\vect{r}+\ln{|\mathbf{K}|} + N\ln{(2\pi)}\big] 
\end{equation}
where $\vect{r}=\vect{y-m}$ is the residual vector, $N$ is the number of (observed) datapoints and $|\mathbf{K}|$ denotes the determinant of the full covariance matrix.
The GP predictions depend on the choice of kernel describing the correlations between the data points. In this work, we use a \emph{squared exponential} (SE) kernel given by
\begin{equation}\label{kernel}
    k(x,x';\sigma_f,\ell_f)=\sigma_f^2\,e^{-(x-x')^2/2\ell_f^2},
\end{equation}

where $\sigma_f$ and $\ell_f$ determine the amplitude and typical length-scale of the correlations, respectively.
These hyperparameters are optimized by maximizing the log-marginal likelihood in \eqref{LML}; we refer to \cite{rasmussen2006gaussian} for a more detailed discussion.\\

In this work, we focus on testing the consistency of the \lcdm\ model, and thus decide to work in residual space where the best-fit (\lcdm) predictions have been subtracted from the data---effectively choosing \lcdm\ as a GP mean function. More specifically, we decide to work in the space of $\mathcal{D}_\ell=\ell(\ell+1)\mathcal{C}_\ell/2\pi$, where the physical (oscillatory) features would be more pronounced. Having a closer look at Eq. \eqref{LML}, it is seen that if the mean function is a good (enough) fit to the data, the first and second (penalty) terms in \eqref{LML} will tend to prefer \emph{no extra-correlations} (\emph{i.e.} $\sigma_f\simeq0$) or \emph{diverging correlation-lengths} ($\ell_f\to\infty$), as encoded in the GP kernel \eqref{kernel}. In the presence of hidden systematics or in the need for a modification of the mean function, however, a finite value for $(\sigma_f,\ell_f)$ might be statistically preferred.
Therefore, inspecting the two-dimensional likelihood profile $\mathcal{L}(\sigma_f,\ell_f)$ can yield valuable information on the model and the dataset under consideration \citep{PhysRevD.87.023520,Aghamousa_2017,Keeley:2019hmw,Krishak:2021fxp,Calderon:2023msm}. Thus, if the likelihood is maximized for $\sigma_f\to0$ (or $\ell_f\to\infty$), the mean function is consistent with the data. On the other hand, any significant detection of $\sigma_f\neq0$ can be interpreted as hints of underlying structures or systematics in the data that cannot be properly accounted for by the model, given by a smooth deformation with a typical amplitude and correlation length given by the preferred values of $(\sigma_f,\ell_f)$.

\begin{table}[ht]
    \centering
    \caption{Uniform prior ranges in the hyperparameters for TT,TE and EE.}
    \label{tab:priors_params}
    {\rowcolors{2}{lgray}{white}
    \begin{tabular}{ccc}
        \hline
		Parameter &  $\log_{10}\sigma_f$ & $\log_{10}\ell_f$  \\ 
		\hline
		TT &  $[-3,2]$ & $[0,4]$  \\ 
		TE  & $[-3,0.5]$ & $[0,4]$   \\ 
		EE  & $[-3,0.5]$ & $[0,4]$  \\ 
		\hline
    \end{tabular}}
\end{table} 
\section{Results and Discussions}

In this section, we confront the best-fit \lcdm\ predictions with the various CMB observations. Extending upon \cite{Aghamousa_2017}, our goal is to test the consistency of the \lcdm\ model, update the analysis to include the most recent CMB measurements described before, namely the \planck\ 2018 from \texttt{Plik}, CamSpec (\texttt{NPIPE}) PR4, ACT DR4 and SPT-3G data releases, and test the consistency between these datasets.\\

\begin{figure}
    \centering
    \includegraphics[scale=0.6]{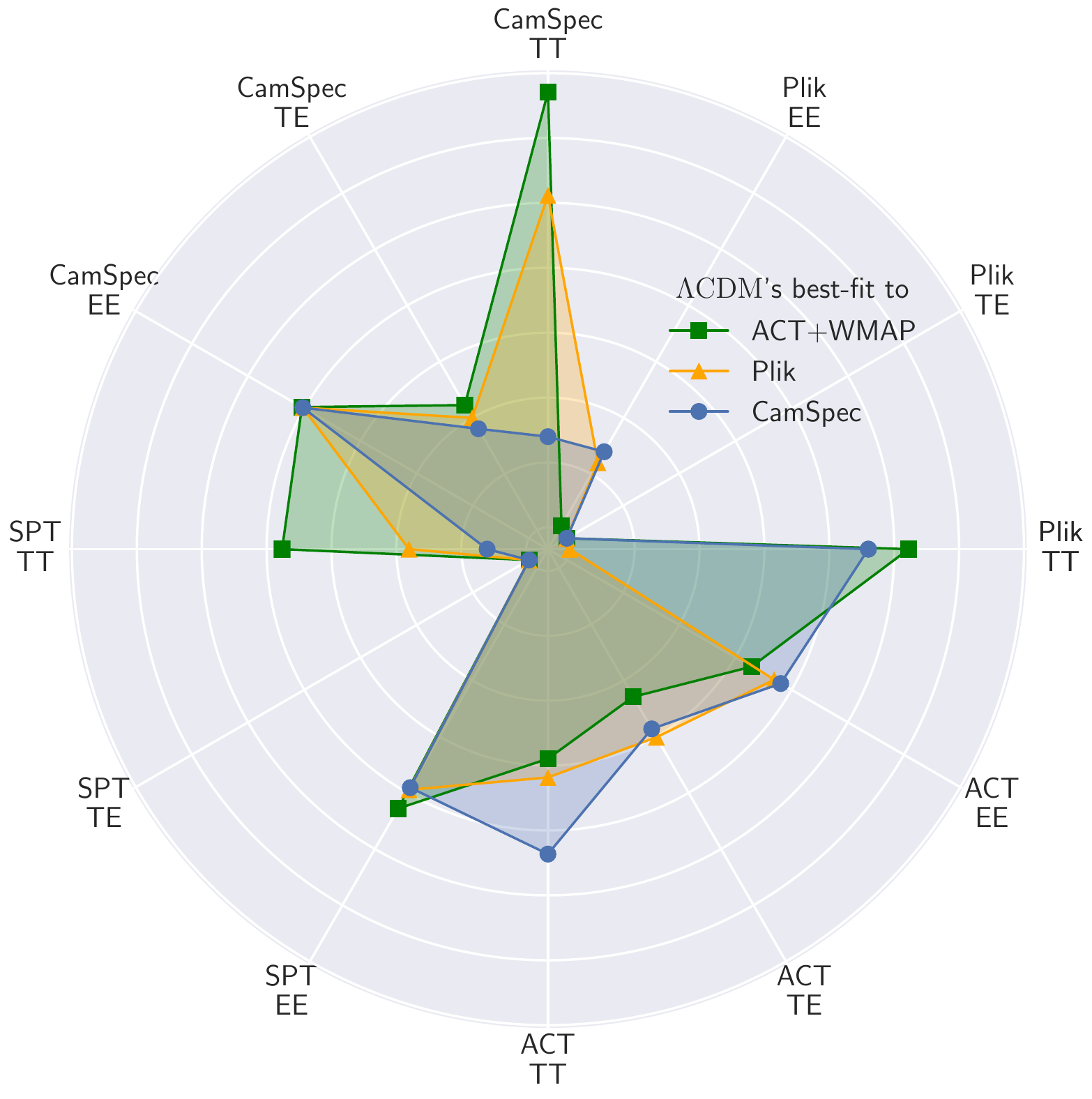}
    \caption{Visual overview of the results of this analysis. The labels in the angular direction correspond to the different measurements used in the GP analysis. The radial direction indicates the level of ``disagreement'' between a given set of measurements and the best-fit predictions that are used as mean functions, shown as green, orange and blue lines. The closer a point lies to the origin, the more in agreement the measurements are with the set of $\Dell$'s from the \lcdm\ mean function. Conversely, points far away from the origin admit large improvements in fit, given by smooth deformations of the corresponding (\lcdm) mean function.}
    \label{fig:Summary}
\end{figure}

Before we delve into the details of the analysis, we start by giving a brief overview of the results. Fig. \ref{fig:Summary} nicely captures the main takeaways of our analysis. The radial distance to the origin translates the level of ``tension''\footnote{The exact metric for quantifying the level of (dis)agreement is not relevant here, but it is related to the number quoted in Fig. \ref{fig:heatmap}; see Section \ref{sec:summary} for a more detailed discussion.} between a given set of measurements and the \lcdm's best-fit predictions. It is clear that the CamSpec and ACT directions shows overall the largest discrepancies, but it is also seen that this is statement is somewhat dependent on the choice of mean function.
To further investigate this, we will thoroughly discuss the results using a single mean function in the main body of the paper.
More specifically, we choose the \lcdm\ best-fit to CamSpec PR4 data as a mean function in our Gaussian Process (i.e. in Eq. \eqref{mean_gp}) since it is obtained from the latest (re)analysis of the Planck data \cite{Efstathiou_2021,Rosenberg:2022sdy}. The CamSpec (\texttt{NPIPE}) PR4 analysis includes differences in the treatment of polarisation, calibration, and systematic corrections. It also contains significantly more sky fraction (more data than the official Planck data release) and yields the tightest constraints in terms of cosmological parameters. In Section \ref{LCDM_Planck}, we will extensively discuss the results using \planck\ measurements, both from the CamSpec and \texttt{Plik} analyses. The results of our analysis using ground-based observations, namely ACT and SPT, will be discussed in 
Section \ref{sec:ACTSPT}. Finally, we comment on the role of the mean function and summarize our results in Section \ref{sec:summary}.  Further details are also given in the Appendix.

\subsection{Consistency of \tpdf{\lcdm } with \planck}\label{LCDM_Planck}

\subsubsection{CamSpec PR4}\label{LCDM_CSPR4}

\begin{figure}[t]
    \centering
    \includegraphics[width=\textwidth]{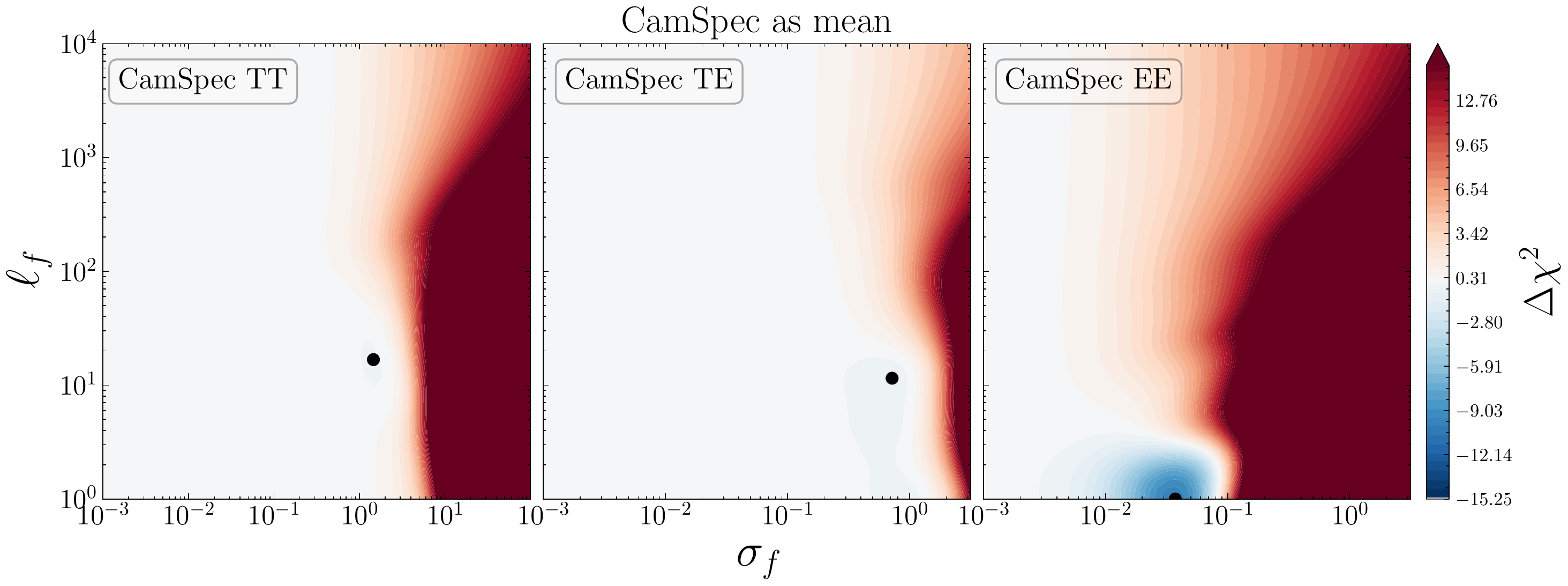}
    \caption{2D posterior distributions as a function of \hyperpars\ for the \emph{CamSpec PR4} data, and \lcdm\ best-fit to the same data as mean function. The color bar shows the improvement in fit, where $\Delta\chisq=-2(\ln\Lkl^{\rm GP}-\ln\Lkl^{\Lambda \rm CDM})$ and $\ln\Lkl^{\rm GP}\hyperpars$ is the log marginal likelihood in Eq. \eqref{LML}.}
    \label{fig:posteriors_CamSpec-CamSpec}
\end{figure}

We start by considering the \lcdm's best fit to the latest CamSpec data \cite{Rosenberg:2022sdy} as the mean function in our analysis. In Fig. \ref{fig:posteriors_CamSpec-CamSpec}, we show the two-dimensional likelihood profiles for the CamSpec residuals with respect to \lcdm's best-fit $\mathcal{D}_\ell$'s, as a function of \hyperpars. The color bar shows the goodness of fit, where $\Delta\chisq=-2(\ln\Lkl^{\rm GP}-\ln\Lkl^{\Lambda \rm CDM})$ and $\ln{\mathcal{L}^{\rm GP}}$ is the log-marginal likelihood (LML), defined in Eq. \eqref{LML}. Negative values of $\Delta\chisq$ (in blue) reflect regions in parameter space yielding an improvement in the fit with respect to \lcdm. Conversely, red-colored regions correspond to deviations from the mean function leading to a degraded fit to the data ($\Delta\chisq>0$), whereas white-shaded regions represent no improvement at all. The black dot represents the set of hyperparameters \hyperpars\ yielding the highest likelihood and we report the corresponding improvement in fit\footnote{Strictly speaking, this quantity is analogous to a \chisq\ difference, but the exact $\Delta\chisq$ values should not be interpreted in the usual sense, as the LML receives additional (possibly large) contributions from the second and third terms in Eq.\eqref{LML} which depend on the values of the kernel hyperparameters \hyperpars.} ($\Delta\chisq$) in Table \ref{tab:deltachi2_CSPR4}. As mentioned before, if the mean function is a good description of the data, the LML in \eqref{LML} should peak at $\sigma_f\to0$ and/or $\ell_f\to\infty$. In other words, no smooth deviations away from the best-fit \lcdm\ are needed to explain the data. 
However, if the LML peaks at finite (possibly large) values of \hyperpars, it might point towards the need for a different mean function or indicate the existence of hidden structures/systematics in the data.\\

In this case, as can be seen from Fig. \ref{fig:posteriors_CamSpec-CamSpec}, the \lcdm\ model provides a relatively good fit to TT, TE, and EE data. The LML seems to prefer small deviations from the best fit \lcdm\ spectra:
$\sigma_f\lesssim1$ for TT and TE and $\sigma_f\lesssim0.1$ for the case of EE. Any larger deviation from the mean function is highly penalized by the data, as can be seen by the color bar on the right ($\Delta\chisq>0$ meaning a degraded fit), and the improvement in fit by the GP is negligible in both TT and TE; see Table \ref{tab:deltachi2_CSPR4}.
Meanwhile, there is a noticeable improvement in fit in EE for $\ell_f\lesssim1$. Such a GP realisation is essentially a white noise where the values at each $\ell$ are uncorrelated with each other. Preferences toward the inclusion of white noise indicate that the fluctuations in the data around the mean are more than what is expected from the data covariance matrix. In other words, the covariance matrix may have been slightly underestimated for this EE data. This may alter the weights and hence the optimality of the cosmological parameter estimation, but is less likely to have a significant effect on the estimated parameter values. \\

In fact, the LML is expected to have a local minimum at some $\sigma_f$ with $\ell_f \ll 1$ about half of the time. Taking the limit $\ell_f \rightarrow 0$ in \eqref{LML}, we found that the necessary and sufficient condition for such a minimum is given by ${\| \Sigma^{-1} \bf{r} \|^2 > \rm{tr}\left( \Sigma^{-1}\right)}$, where $\bf{r}$ is the residual vector. Assuming that the mean and the covariance matrix are exact, the expected value of the left-hand side is equal to the right-hand side. When the residual vector is large enough either by statistical fluctuations and/or underestimation of the errors, then we expect to see a local minimum at some $\sigma_f$ satisfying ${\| (\sigma_f^2 I + \Sigma)^{-1} {\bf r} \|^2 = {\rm tr}\left[ (\sigma_f^2 I + \Sigma)^{-1}\right]}.$ 
\begin{figure}[h]
    \centering
    \includegraphics[width=0.8\textwidth]{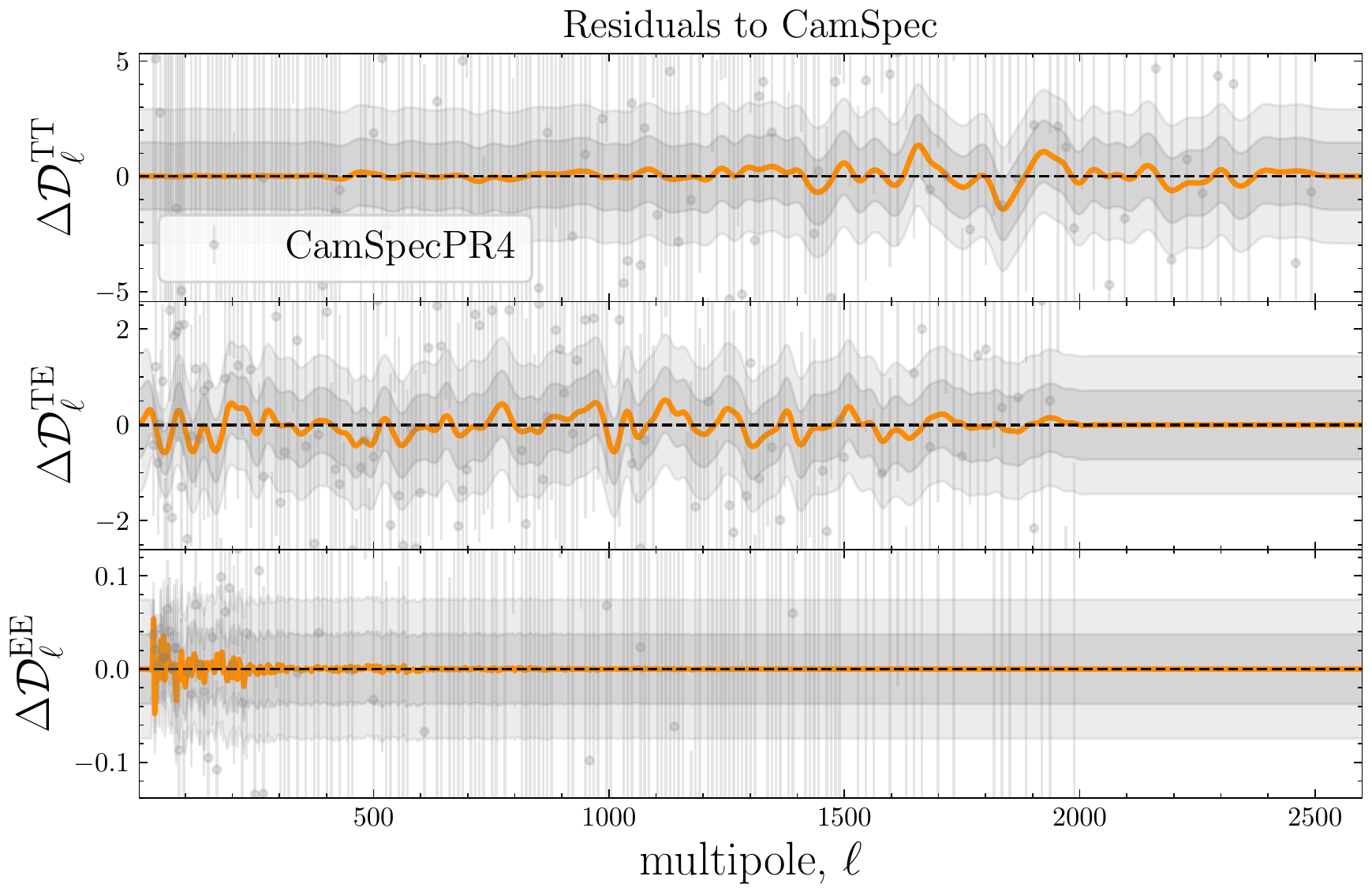}
    \caption{GP reconstructions of $\Delta\Dell\equiv\Dell^{\rm GP}-\Dell^{\Lambda\rm CDM}$ for the set of hyperparameters maximizing the log-marginal likelihood in Eq. \eqref{LML} when using CamSpec PR4 data and the corresponding \lcdm\ best-fit spectra as mean functions. The solid line and shaded regions correspond to the mean and $2\sigma$ confidence intervals around it, respectively.}
    \label{fig:GP_CamSpec}
\end{figure}\\
Despite this fact, it is still intriguing that we find $\Delta\chi^2$ as low as $-10.21$ in EE for CamSpec PR4. As we discuss in Appendix \ref{sec:HvsL}, most of these improvements in fit come from low-$\ell$ data, and notably is \textit{not} present in the \planck\ Official Release (PR3). Our non-parametric approach using GP indicates that the updated analysis pipeline of CamSpec PR4 has caused the residuals in the EE coadded spectrum to vary more than the expected amount given by the covariance matrix; the errors seem to have been slightly underestimated. Indeed, we confirm that the usual chi-squared statistic for the EE data in the range $30\le\ell\le650$ is $\chisq=709$, larger than the expected value of $620$ by $2.51\sigma$.\footnote{For the full data range of $30\le\ell\le2000$, $\chisq=2023$, which is $0.83\sigma$ above the expected value of 1971. Note that here we look at the TT+TE+EE best-fit values.} This excess in residual error has been investigated in \citep[][see e.g. Table 1]{Rosenberg:2022sdy} and is reported to be statistically consistent with random fluctuations. Moving one step further than these simple chi-squared checks, we investigate the scales where these excess residuals are correlated through GP analyses. The fact that the marginal likelihood is maximised at small $\ell_f$ implies that these excess variances appear to be uncorrelated random fluctuations rather than smooth deformations in the mean, unlike the cases of TT and TE. Nonetheless, the deviations from zero are $\mathcal{O}(10^{-2})$, with relatively minor improvements in fit, suggesting \lcdm\ is a good description of the \planck\ data.\\

We then use the set of \hyperpars\ maximizing the likelihood in Eq. \eqref{LML} (shown as a black dot in Fig. \ref{fig:posteriors_CamSpec-CamSpec}) to obtain the mean (in orange), $68\%$ and $95\%$ C.L. (gray-shaded bands) shown in Fig. \ref{fig:GP_CamSpec}, using Eqs. \eqref{mean_gp} and \eqref{cov_gp}, respectively. Again, we see that the reconstructions are perfectly consistent with zero across the entire multipole range covered by the data.\\

These results suggest that despite the aforementioned issues in EE, the best-fit \lcdm\ model is overall consistent with the CamSpec data. This should not come as a surprise, since we have chosen the best-fit predictions to CamSpec data as the mean function in our analysis. However, as explained before, this serves as a consistency test for the different CMB measurements. In the presence of systematics, or physics beyond \lcdm, some inconsistencies might appear when using a (possibly incorrect) \lcdm\ mean function.

\begin{table}
    \centering  
    \caption{$\Delta\chisq=-2(\ln\Lkl^{\rm GP}-\ln\Lkl^{\Lambda \rm CDM})$ improvements in fit for the ground and space-based experiments obtained with the GP using \lcdm's best-fit to CamSpec PR4 as mean function.}
    {\rowcolors{2}{lgray}{white}
    \begin{tabular}{cccccc}
        \hline
		$\Delta\chisq$ & \planck\ 2018 & CamSpec PR4  & ACT DR4 & SPT-3G \\ 
		\hline
		TT & $-19.94$ & $-0.35$ & $-15.11$ & $-0.124$    \\ 
		TE & $-0.001$ &$-0.75$ & $ -2.62$ & $-0.002$  \\ 
		EE & $-0.446$ & $-10.21$& $-7.89$ & $-8.916$  \\ 
		\hline
    \end{tabular}}
    \centering
    \label{tab:deltachi2_CSPR4}
\end{table}

\subsubsection{Planck 2018 - Baseline/Official Release}\label{P18-CamSpec}

\begin{figure}[h]
    \centering
    \includegraphics[width=\textwidth]{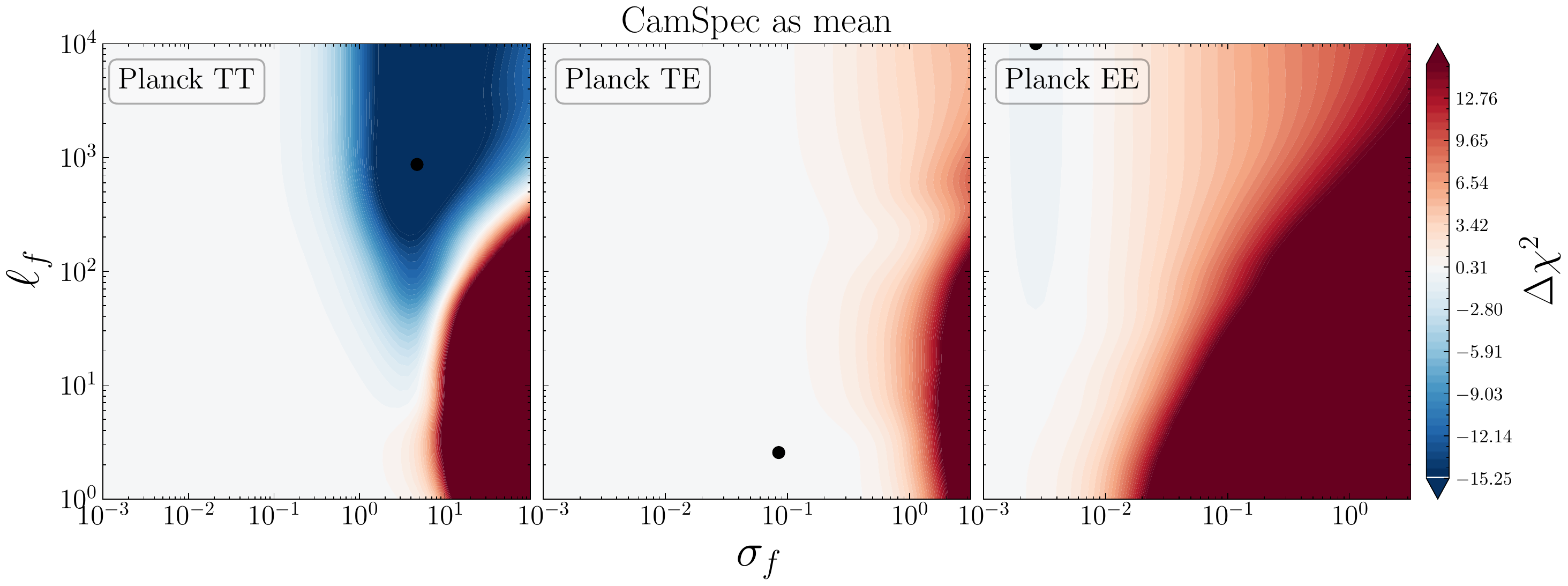}
    \caption{2D posterior distributions as a function of \hyperpars\ for the \emph{Planck 2018} data, and \lcdm\ best-fit to the \emph{CamSpec PR4} data as mean function. The color bar shows the improvement in fit, where $\Delta\chisq=-2(\ln\Lkl^{\rm GP}-\ln\Lkl^{\Lambda \rm CDM})$ and $\ln\Lkl^{\rm GP}\hyperpars$ is the log marginal likelihood in Eq. \eqref{LML}.}
    \label{fig:posteriors_CamSpec-P18}
\end{figure}

Similarly, we take the \lcdm's best-fit predictions to CamSpec data and confront it to the official \planck\ likelihood (\texttt{Plik}) as a way of testing the internal consistency of the \planck\ measurements and the differences between the CamSpec and \texttt{Plik} likelihoods. In Fig. \ref{fig:posteriors_CamSpec-P18}, we show the 2D LML profiles for TT, TE, and EE measurements from \texttt{Plik} (which we simply refer to as Planck). As discussed before, finding a significantly low $\Delta\chi^2$ by our GP means that a smooth deformation away from the mean function is preferred, which would hint towards inconsistencies between CamSpec and \texttt{Plik} in this case.
While we find TE and EE to be consistent, we find notable inconsistencies in TT for the two likelihoods.
There is a large improvement in fit from GP for $\sigma_f\simeq5$ and $\ell_f\simeq10^3$; corresponding to the blue regions in Fig. \ref{fig:posteriors_CamSpec-P18}.

\begin{figure}[h]
    \centering
    \includegraphics[width=0.8\textwidth]{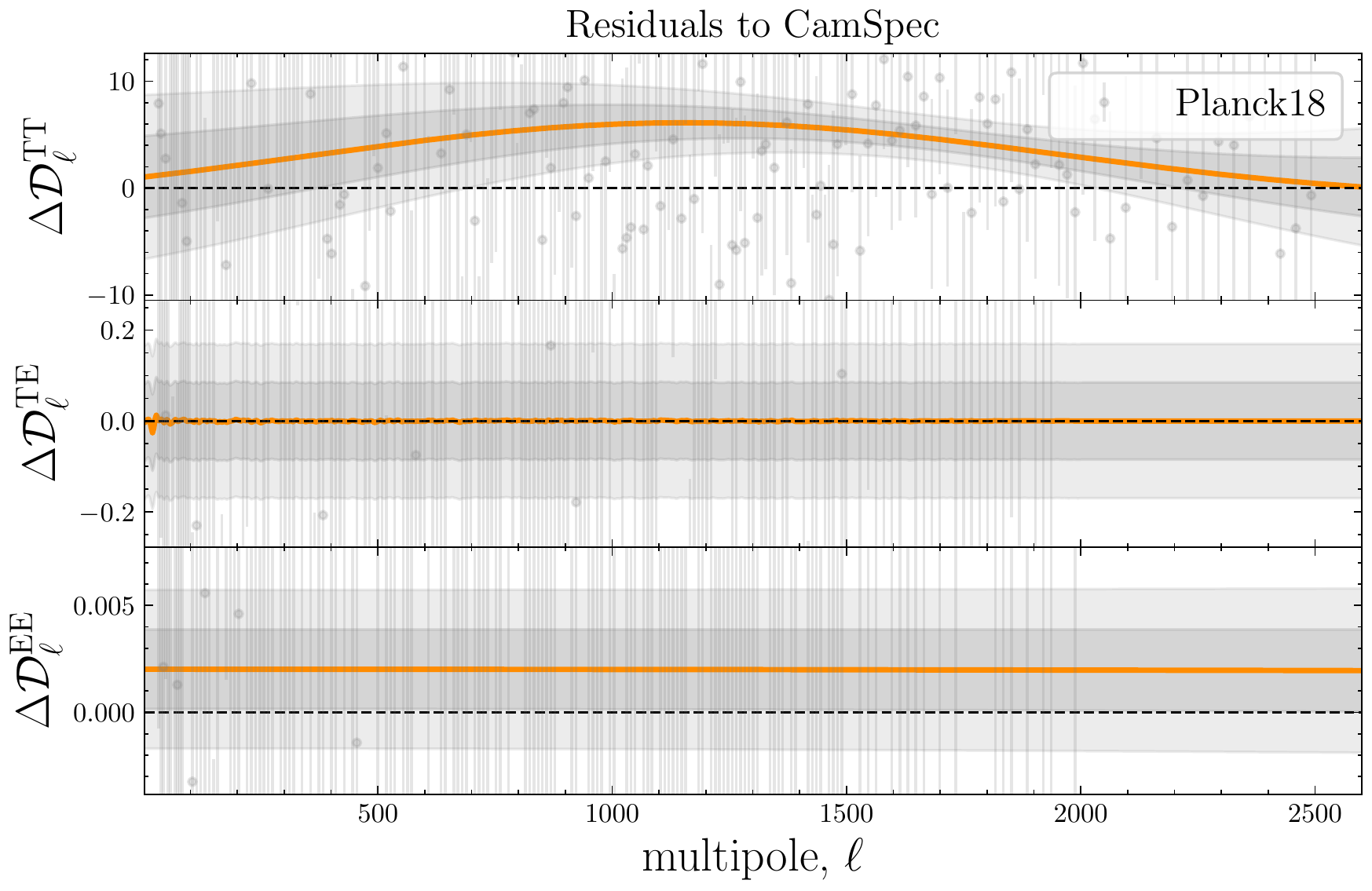}
    \caption{GP reconstructions of $\Delta\Dell\equiv\Dell^{\rm GP}-\Dell^{\Lambda\rm CDM}$ for the set of hyperparameters maximizing the log-marginal likelihood in Eq. \eqref{LML} when using \planck\ data and the \lcdm\ best-fit spectra to CamSpec as mean functions. The solid line and shaded regions correspond to the mean and $2\sigma$ confidence intervals around it, respectively.}
    \label{fig:GP_P18-CamSpec}
\end{figure}

As seen in the upper panel of Fig. \ref{fig:GP_P18-CamSpec}, the corresponding GP reconstruction also shows large deviations ($>3\sigma$) from zero.
The corresponding improvements in fit, $\Delta\chisq$, are reported in the first column of Table \ref{tab:deltachi2_CSPR4}. See also Table 3 in \cite{Rosenberg:2022sdy} and discussions therein for a more detailed comparison between the official Planck release (PR3) and CamSpec (PR4) within the \lcdm\ model.

Interestingly, our GP analysis shows that these smooth deformations away from the CamSpec mean function not only fit the \texttt{Plik} likelihood significantly better but also have a similar correlation length $\ell_f$ to those that improve the fit to the ACT data (Figs. \ref{fig:posteriors_ACT-CamSpec} and \ref{fig:GP_ACT}), albeit with a smaller amplitude ($\sigma_f$); see Section \ref{ACTDR4} below. These results imply that the best-fit predictions from CamSpec and \texttt{Plik} differ mainly in the TT power spectrum.

The fact that the GP captures differences in temperature between CamSpec and Plik is not so surprising. The temperature measurements from the official \planck\ data release are known to prefer deviations away from the standard model in some simple (one-parameter) extensions to \lcdm. Two well-known examples are the tendencies of the TT spectra to favour an excess of lensing amplitude $A_{L}>1$  -- which should be equal to unity in \lcdm\ -- and the preference for a closed Universe, with positive spatial curvature $\Omega_{k,0}<0$. As already noted in \cite{Planck:2015fie}, these tendencies in TT to favour higher lensing amplitudes have similar effects in the $\Cell$'s as other parameters, such as curvature or the dark energy equation of state in extended models.
Interestingly, these ``anomalies'' disappear when going from \texttt{Plik} to CamSpec (i.e. $A_{L}\to1$ and $\Omega_{k,0}\to0$) and are attributed to statistical fluctuations in the multipole range $800<\ell<1600$ \cite{Efstathiou_2021}. Our GP reconstruction for the TT power spectrum also shows significant deviations from zero in the same multipole range; see the upper panel of Fig. \ref{fig:GP_P18-CamSpec}. However, we note that our GP analysis prefers larger correlation scales ($\ell_f\simeq10^3$) than what one would expect from random fluctuations, even considering the off-diagonal correlations from lensing.\\

Finally, we should stress that in this comparison, we use the best-fit power spectra to the CamSpec likelihood. An alternative analysis is to study deviations away from the TTTEEE best-fit to the official \planck\ data release, shown in green in Fig. \ref{fig:residuals}. 
We refer the reader to Appendix \ref{sec:P18mean}, where we use the best-fit $\Dell$'s to \texttt{Plik} instead, and test the consistency with both CamSpec measurements and other ground-based experiments.

\subsection{Consistency of  \planck\ with ground-based experiments (within \tpdf{\lcdm})}\label{sec:ACTSPT}

Next, we use the same mean function as before and look for potential structures in the residuals of ACT and SPT data. If \lcdm\ is the correct model describing the CMB anisotropies up to $\ell\simeq4000$, and their parameters are accurately estimated by the CamSpec analysis, then the two-dimensional distributions of the hyperparameters \hyperpars\ should not prefer any finite, non-vanishing values. The presence of unaccounted systematics, or discrepancies between the experiments, would however be reflected in the two-dimensional likelihood profiles if there are any.

\subsubsection{ACT DR4}\label{ACTDR4}

\begin{figure}[h]
    \centering
    \includegraphics[width=\textwidth]{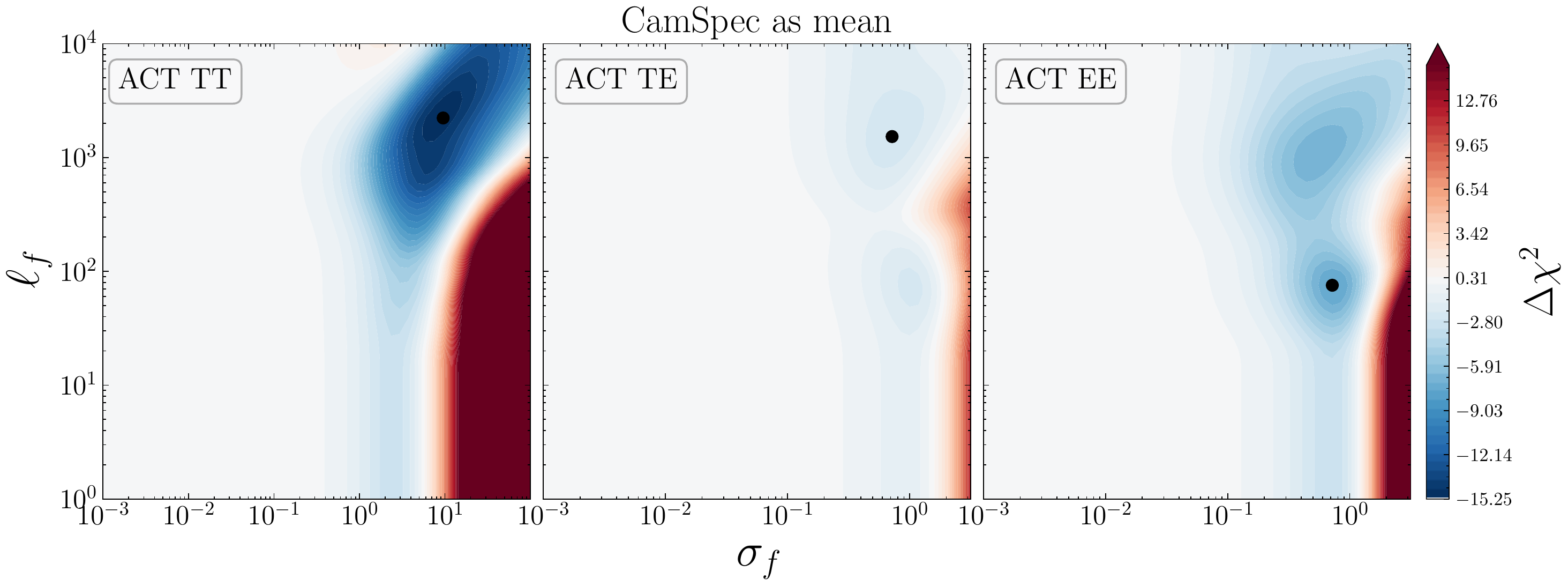}
    \caption{2D posterior distributions as a function of \hyperpars\ for the ACT DR4 data, and \lcdm\ best-fit to CamSpecPR4 data as mean function. The color bar shows the improvement in fit, where $\Delta\chisq=-2(\ln\Lkl^{\rm GP}-\ln\Lkl^{\Lambda \rm CDM})$ and $\ln\Lkl^{\rm GP}\hyperpars$ is the log marginal likelihood in Eq. \eqref{LML}.}
    \label{fig:posteriors_ACT-CamSpec}
\end{figure}

In Fig. \ref{fig:posteriors_ACT-CamSpec} we show the posteriors for \hyperpars\ when using ACT DR4 data and \lcdm's best-fit to CamSpec as mean function. In this case, an interesting feature appears in the TT data. The LML peaks at $\hyperpars \simeq(9,2\times10^3)$ where the GP finds an improvement in fit with respect to the mean function (\lcdm), corresponding to a $\Delta\chisq=-15.11$; see Table \ref{tab:deltachi2_CSPR4}. Interestingly, the TE and EE posteriors show similar (bimodal) distributions, with a preference for non-vanishing values of \hyperpars, although the statistical significance of these deviations from \lcdm\ is milder than in the TT case. The improvements in fit are reported in the third column of Table \ref{tab:deltachi2_CSPR4}.

\begin{figure}[h]
    \centering
    \includegraphics[width=0.8\columnwidth]{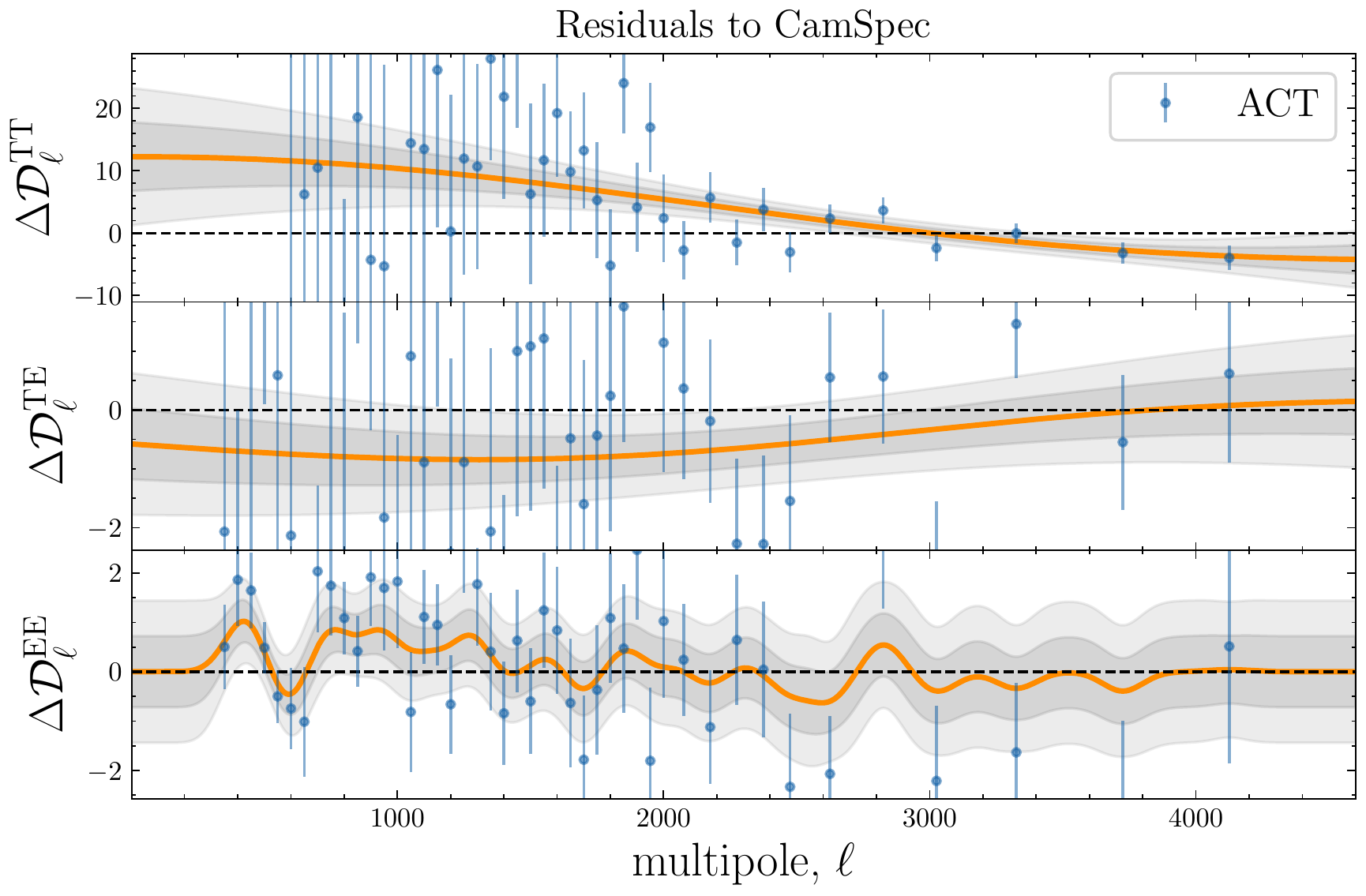}
    \caption{GP reconstructions of $\Delta\Dell\equiv\Dell^{\rm GP}-\Dell^{\Lambda\rm CDM}$ for the set of hyperparameters maximizing the log-marginal likelihood in Eq. \eqref{LML} when using ACT DR4 data and  \lcdm\ best-fit to CamSpec PR4 as mean function. The solid line and shaded regions correspond to the mean and $2\sigma$ confidence intervals around it, respectively. Note that the multipole range is extended with respect to Fig. \ref{fig:GP_P18-CamSpec}.}
    \label{fig:GP_ACT}
\end{figure}

In Fig. \ref{fig:GP_ACT}, we show the mean and $2\sigma$ reconstructions from the GP when using the set of hyperparameters maximizing the LML in Eq. \eqref{LML}, shown as black dots in Fig. \ref{fig:posteriors_ACT-CamSpec}. Note that the ACT reconstructions of the TT spectra seem to prefer lower amplitudes at $\ell\lesssim3000$ (at more than $2\sigma$) and a slightly larger amplitude at $\ell\gtrsim3000$ with respect to what is predicted by \lcdm's best-fit to the CamSpec data. This is yet another (non-parametric) indication that the ACT data seem to favor a scale-invariant, Harrison-Zel'dovich ($n_s\simeq1$) spectrum of fluctuations \citep{PhysRevD.103.063529,https://doi.org/10.48550/arxiv.2210.06125,Corona:2021qxl,Giare:2022rvg}. At the same time, a larger value for $n_s$ might also imply an increased value of $H_0$\footnote{However, we should note that such a shift in the cosmological parameters, $H_0\to73\;\rm km/s/Mpc$ and $n_s\to1$, would typically lead to larger values of $S_8$, worsening the fit to low-redshift (weak-lensing) measurements of the clustering amplitude.}, through a reduction of the size of the sound horizon \citep{Ye_2021,Jiang_2022}.\\

While the LML improvements at $\ell_f \sim 2000 $ relate to the overall scaling through $n_\mathrm{s}$, the other mode in LML found at $\ell_f \sim 90$ in TE and EE spectra may be closely related to the cosmological parameters affecting the width and height of the acoustic peaks. Roughly speaking, a realisation of a GP with $\ell_f \sim 90$ has a typical full width at half maximum (FWHM) of $\sim 210$ and is likely to have oscillations that mimic the acoustic peaks in the CMB ($\Delta\ell \sim 300$). Indeed, the GP reconstruction of Fig. \ref{fig:GP_ACT} for EE spectra has oscillation scales similar to those of the differences in the best-fit predictions from \actpwmap\ and CamSpec data (red line in Fig. \ref{fig:residuals}). The features present in our non-parametric reconstructions can therefore be a manifestation of the mild discordance in the ($\omega_b,\omega_c$)-plane between Planck and ACT (seen for instance in Fig. 7 in \cite{SPT-3G:2022hvq}), or vice versa.
An interesting feature is also captured in the EE reconstructions around $\ell\sim 400-700$, with milder oscillations extending up to $\ell\sim3000$. These might be a slight hint of new features, a manifestation of the mild disagreement in the estimated cosmological parameters, or some unaccounted systematics affecting the low/high-$\ell$ part of the ACT and/or \planck\ data. Together with the issue in TT, it might explain why recent analyses reach slightly different conclusions when considering ACT data alone, or performing cuts at a given $\ell_{\rm max}$ in the \planck\ data \citep[e.g.][]{Poulin:2021bjr,ACT_EDE,Poulin:2023lkg}. Our results seem to support previous findings in the context of \lcdm\ and simple extensions \cite[e.g.][]{PhysRevD.103.063529,Galli_2022,Corona:2021qxl,DiValentino:2022rdg}, suggesting that these mild discrepancies are mainly driven by the ACT data and in particular by the TT measurements.
Whether such discrepancies arise from physical, systematic, or statistical origin, however, remains to be determined by upcoming (more precise) CMB observations. In particular, the ACT collaboration is soon expected to update its results with the ACT DR6 data release.

\subsubsection{SPT-3G}

\begin{figure}[h]
    \centering
    \includegraphics[width=\textwidth]{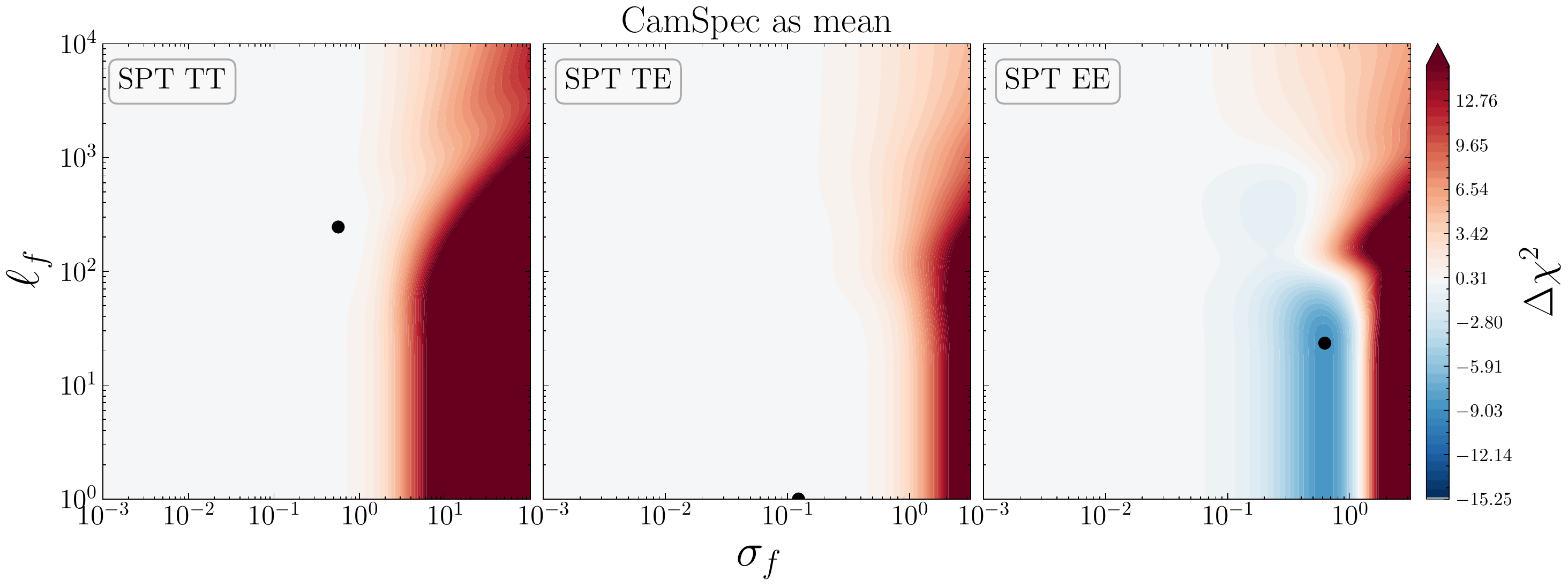}
    \caption{2D posterior distributions as a function of \hyperpars\ for the SPT-3G data, and \lcdm\ best-fit to CamSpec data as mean function. The color bar shows the improvement in fit, where $\Delta\chisq=-2(\ln\Lkl^{\rm GP}-\ln\Lkl^{\Lambda \rm CDM})$ and $\ln\Lkl^{\rm GP}\hyperpars$ is the log marginal likelihood in Eq. \eqref{LML}.}
    \label{fig:Posteriors_SPT-CS}
\end{figure}
\begin{figure}[h]
    \centering
    \includegraphics[width=0.8\columnwidth]{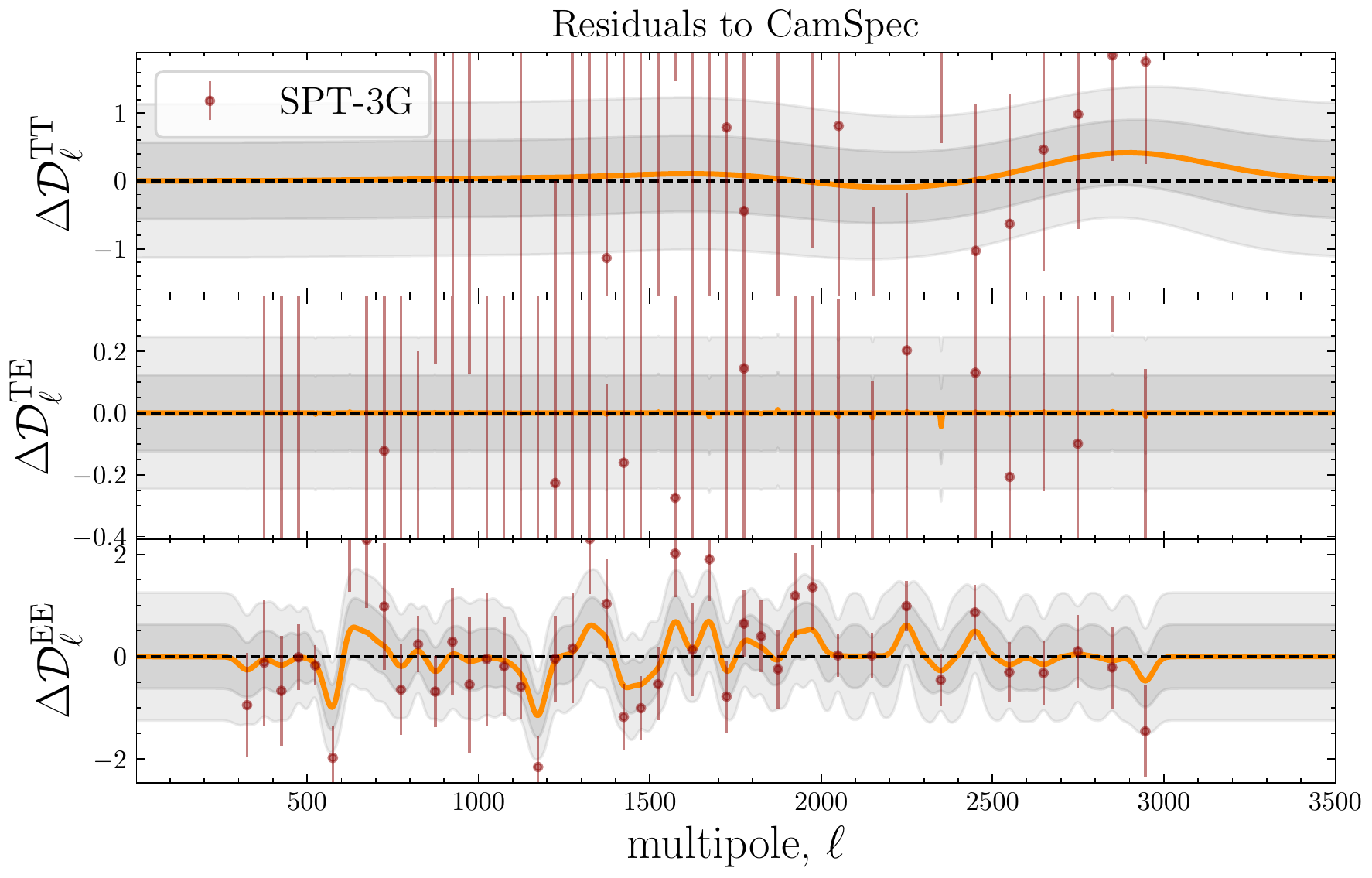}
    \caption{GP reconstructions of $\Delta\Dell\equiv\Dell^{\rm GP}-\Dell^{\Lambda\rm CDM}$ using the set of hyperparameters maximizing the log-marginal likelihood in Eq. \eqref{LML} when using SPT-3G data and  \lcdm\ best-fit to CamSpec PR4 as mean function. Solid line and shaded regions correspond to the mean and $2\sigma$ confidence intervals around it, respectively. Note that the multipole range is extended with respect to Fig. \ref{fig:GP_P18-CamSpec}}
    \label{fig:GP_SPT}
\end{figure}
Similarly, we inspect for structures in the residuals of SPT-3G data, when subtracting \lcdm's best fit to CamSpecPR4 data. The results are shown in Fig. \ref{fig:Posteriors_SPT-CS}.
The posteriors of the temperature auto-correlation (TT) and temperature/polarisation cross-correlation (TE) are in good agreement with the \lcdm\ predictions, and the GP finds negligible improvements with respect to \lcdm; see also Fig. \ref{fig:GP_SPT}. However, this could also be explained by the larger uncertainties in the temperature measurements with respect to \planck\ (see Fig. \ref{fig:residuals}). The situation is slightly different for EE, suggesting again a bimodal distribution in the \hyperpars-plane, with typical deviations from a zero mean-function of the order $\sigma_f\lesssim1$ and with a preferred correlation length of $\ell_f\simeq30$, indicating that the improvement in fit is likely due to subtle oscillations around the \lcdm\ best-fit predictions. The improvements in fit with respect to \lcdm\ are again reported in the last column of Table \ref{tab:deltachi2_CSPR4}, with a maximum $\Delta\chi^2=-8.196$ for EE, which is of the same order of magnitude as the improvement in fit as for the case of ACT data. Curiously, the SPT reconstructions also show a prominent feature at intermediate scales ($\ell\sim500-700$). As discussed before, these oscillations might be linked to the mild differences in the cosmological parameters (such as $\omega_b$ or $\omega_c$) affecting the width, height and position of the acoustic peaks with respect to the ones inferred by \planck. We should mention that the SPT-3G results are overall consistent with those of \planck\ at the parameter level; see Table IV and Fig. 7 in \cite{SPT-3G:2022hvq}. However, our results suggest that the very mild differences between the two are mostly driven by the EE measurements; in agreement with the conclusions from Fig. 3 in \cite{SPT-3G:2022hvq}.

\subsection{Summary of the results}\label{sec:summary}

\begin{figure}[h]
    \centering
    \includegraphics[width=\textwidth]{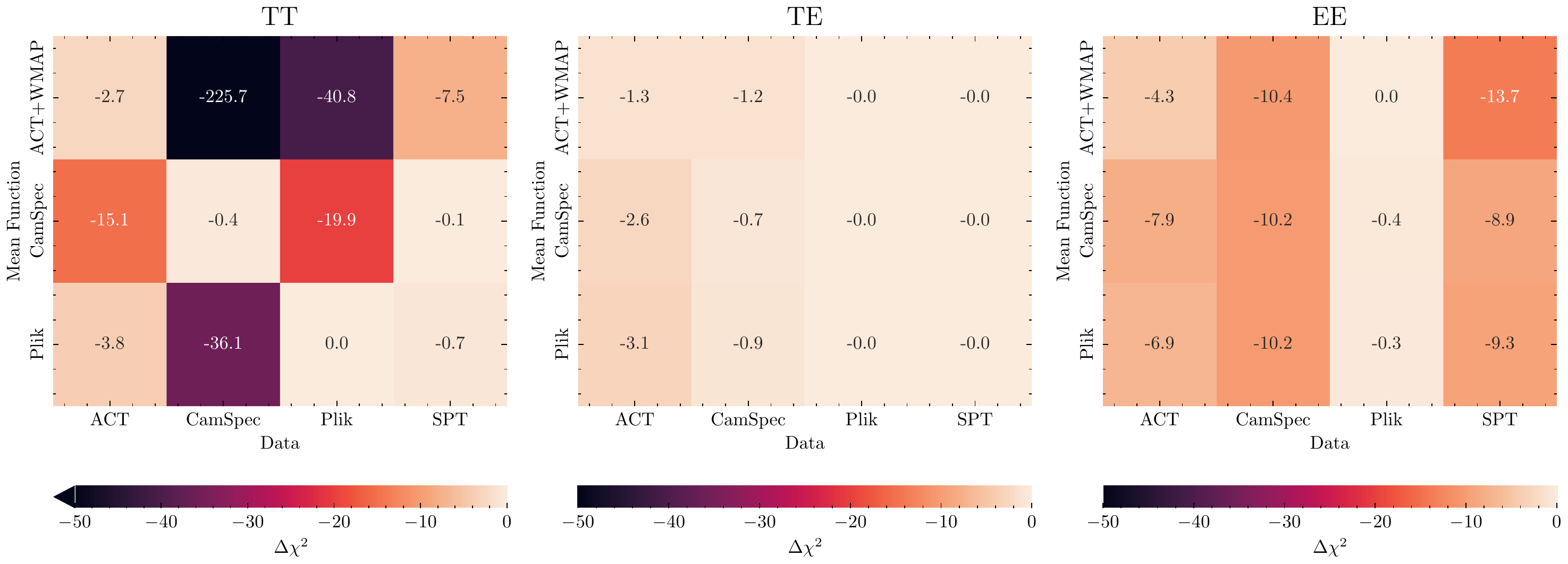}
    \caption{Summary plot assessing the consistency/discrepancy between the different choices of mean function and the different datasets. The row indicates the choice of mean function $\vect{m}$, and the column corresponds to the choice of data with serves as ``training points'' $\vect{y}$ for the GP, Eq. \eqref{LML}.
    We remind the reader that the quantity $\Delta\chisq=-2(\ln\Lkl^{\rm GP}-\ln\Lkl^{\Lambda \rm CDM})$ is analogous to a $\chisq$, but the exact values should be interpreted with caution. It is only meant to represent a statistical preference for a particular GP-deformation of the considered mean function.}
    \label{fig:heatmap}
\end{figure}

In the previous sections, we tested the consistency of the 
CMB measurements assuming \lcdm's best-fit predictions ($\Dell^{XY}$) to CamSpec data. However, as understood from Eqs. \eqref{mean_gp} and \eqref{LML}, the assumed choice of mean function $\vect{m}$ plays a crucial role in our analysis. In the Appendix, we repeat our analysis using different best-fit (\lcdm) models as mean functions; see also Table \ref{tab:mean_params} for the differences at the parameter level. 
In this section, we aim to summarize some of the key results of our analyses using different mean functions, which correspond to the different rows in Fig. \ref{fig:heatmap}. We refer the reader to the Appendix for more figures and detailed discussions on these results.

\begin{itemize}
    \item We find mild inconsistencies in the temperature auto-spectrum between CamSpec and \texttt{Plik}. As discussed in Section \ref{P18-CamSpec}, the temperature spectra from the official \planck\ 2018 (PR3) release are known to show preferences for a closed Universe or an excess of lensing. In the base \lcdm\ model, these parameters are expected to be $\Omega_{k,0}=0$ and $A_L=1$, respectively. These issues were reported and discussed already in the first \planck\ 2013 results \cite{Planck:2013pxb}; see also the follow-up discussions in \cite{Planck:2015fie,P18Cosmo2020}. However, these tendencies in TT disappear in the CamSpec (re)analysis, and are attributed to random statistical fluctuations in the multipole range $800<\ell<1600$ \cite{Efstathiou_2021}. Our non-parametric analysis captures those differences between \texttt{Plik} and CamSpec in the same $\ell$-range (see Fig. \ref{fig:GP_P18-CamSpec}) and is reflected as improvements in fit ($\Delta\chisq$) by our GP, shown in Fig. \ref{fig:heatmap}.
    \item The $\Dell^{\rm TT}$ reconstructed using both Planck data from \texttt{Plik} and ACT data shows noticeable discrepancies (at more than $2\sigma$ level) with \lcdm's best-fit predictions to CamSpec (see Figs. \ref{fig:GP_P18-CamSpec} and \ref{fig:GP_ACT}), with preferences for large correlation lengths and less power at small/intermediate angular scales with respect to the best-fit predictions, thus mimicking a change in the spectral index $n_s\to 1$. However, the TT data from SPT is consistent with both  the best-fit predictions to CamSpec and \texttt{Plik}; see the leftmost panel in Fig. \ref{fig:heatmap} (see also Fig. \ref{fig:Summary}).
    \item The two-dimensional likelihood profiles in EE and posterior distribution of $
    \Delta\Dell^{\rm EE}$ for the CamSpec data seems to be the same, irrespective of the chosen mean function: the GP always seem to find an improvement in fit ($\Delta\chisq\sim10.2$) by adding some uncorrelated noise ($\ell_f\to1$), which could potentially be reflecting a slight underestimation of the covariance in EE; see the CamSpec column for EE in Fig. \ref{fig:heatmap}.
    \item The marginal-likelihood profiles $\mathcal{L}\hyperpars$ for both ACT and SPT are also stable under changes in the mean function. In particular, the reconstructed (oscillatory) features in EE are always present (see Figs. \ref{fig:GP_ACT}, \ref{fig:GP_SPT}, \ref{fig:GP_residuals-P18} and \ref{fig:GP_residuals-ACT+WMAP}), which might be pointing towards a common physical (possibly primordial) origin, new physics beyond the \lcdm\ model, or differences in the inferred cosmological parameters related the width and height of the acoustic peaks; see also the discussion in Section \ref{ACTDR4}.
    \item The \actpwmap\ best-fit predictions within \lcdm\ seem to be in tension ($\gg2\sigma$) with both CamSpec and \texttt{Plik} TT measurements; see Fig. \ref{fig:GP_residuals-ACT+WMAP}. A similar behaviour is observed for SPT-3G TT measurements, although the statistical significance is milder. We stress that such a test is performed merely for the sake of completeness, to effectively replace Planck's large-scale constraints by WMAP's. Statistically, it is not correct to compare the best-fit obtained from \actpwmap\ and compare it to a dataset with much more constraining power, such as \planck/CamSpec.
\end{itemize}

From all these considerations, as is nicely illustrated in Fig. \ref{fig:Summary}, we conclude that the mean function that better describes the different data sets seems to be the \lcdm\ best-fit from the \texttt{Plik} analysis, being perfectly consistent with itself, and only mildly discrepant with ACT and SPT (with the largest discrepancies seen in EE and relatively minor differences in TT). However, as mentioned before, the CamSpec TT measurements seem to be in tension with \lcdm's best-fit to \texttt{Plik}, which might be non-parametric manifestations of the well-known $A_{L}$ and $\Omega_{k,0}$ anomalies present in the firsts (official) \planck\ data releases \cite{Planck:2013pxb,Planck:2015fie,P18Cosmo2020}, but which are no longer present in the CamSpec re-analysis \cite{Efstathiou_2021,Rosenberg:2022sdy}. We should also note that the CamSpec EE data seem to require the addition of uncorrelated noise, irrespective of the chosen mean function.\\

At this stage, the mild disagreement between the experiments is not statistically significant ($\lesssim2\sigma$) to confidently claim a discrepancy, although it is interesting to see that both ACT and SPT seem to prefer additional features in EE with respect to the best-fit \lcdm\ predictions, which could be pointing towards a common physical origin. Note that with the arrival of upcoming CMB surveys such as Simons Observatory, CMB-S4, LiteBird and others, the situation might soon improve and might even shed light on the origin of these mild differences.

\section{Conclusion} \label{Conclusion}

In this work, we used Gaussian Processes (GP) to test the consistency of \lcdm\ with the most-recent CMB observations. In particular, we tested the robustness of the \lcdm\ predictions against space-based observations from the Planck satellite -- both using the latest CamSpec and the final \planck\ data release (official) from the \texttt{Plik} likelihood \citep{Planck_results_2020,Rosenberg:2022sdy,Efstathiou_2021} --  as well as with other ground-based temperature and polarisation measurements by the Atacama Cosmology Telescope (ACT) \citep{Aiola_2020,Choi_2020} and South Pole Telescope (SPT-3G) \citep{Dutcher_2021,SPT-3G:2022hvq} collaborations. We find a mild but noticeable inconsistency between the best fit \lcdm\ predictions from the CamSpec analysis with \planck\ and ACT data, mainly seen in the TT spectra, where the GP finds a non-negligible improvement in fit, with indications for a Harrison-Zel'dovich spectrum with $n_s\simeq1$. This is a non-parametric confirmation of previous results, which supports the idea that the ACT data seem to favor a scale-invariant primordial power spectrum. Our non-parametric analysis also suggests that the covariance in the polarisation (EE) measurements in the CamSpec likelihood might have been slightly underestimated.
Additionally, the EE measurements from both ACT and SPT seem to require additional features at intermediate scales ($\ell \sim 400-700$), extending up to $\ell\sim2500$, which might be pointing towards a common physical (possibly primordial) origin, differences in the inferred cosmological parameters or (less-likely) minor unaccounted-for systematics in the data.\\
Overall, 
our analysis again confirms the robustness of the \lcdm\ predictions when confronted with state-of-the-art CMB measurements; although our analysis shows some interesting features, which could be hints of new physics. The arrival of upcoming CMB experiments such as the Simons Observatory \cite{SimonsObservatory:2018koc}, CMB-S4 \cite{Abazajian:2019eic} and others, will allow for further, more careful exploration of these issues. The method discussed in this work can be readily applied to the upcoming data; hopefully determining whether the mild discrepancies reported here are actually coming from physical, systematic or statistical origin.

\section*{Acknowledgements}

The authors would like to thank Erik Rosenberg for providing us with the latest CamSpec likelihood used in this analysis. RC would like to thank Adrien La Posta for useful discussions. We also thank the anonymous referee for suggesting a comparison between the official \planck\ and CamSpec results.
This work was supported by the high-performance computing cluster Seondeok at the Korea Astronomy and Space Science Institute. DKH would like to acknowledge the support from CEFIPRA grant no. 6704-1. AS would like to acknowledge
the support by National Research Foundation of Korea
NRF2021M3F7A1082053, and the support of the Korea
Institute for Advanced Study (KIAS) grant funded by the
government of Korea.

\appendix
\section{Different mean functions}

\subsection{Best fit to Planck 2018}\label{sec:P18mean}

\begin{figure}[h]
    \centering
    \includegraphics[width=0.475\columnwidth]{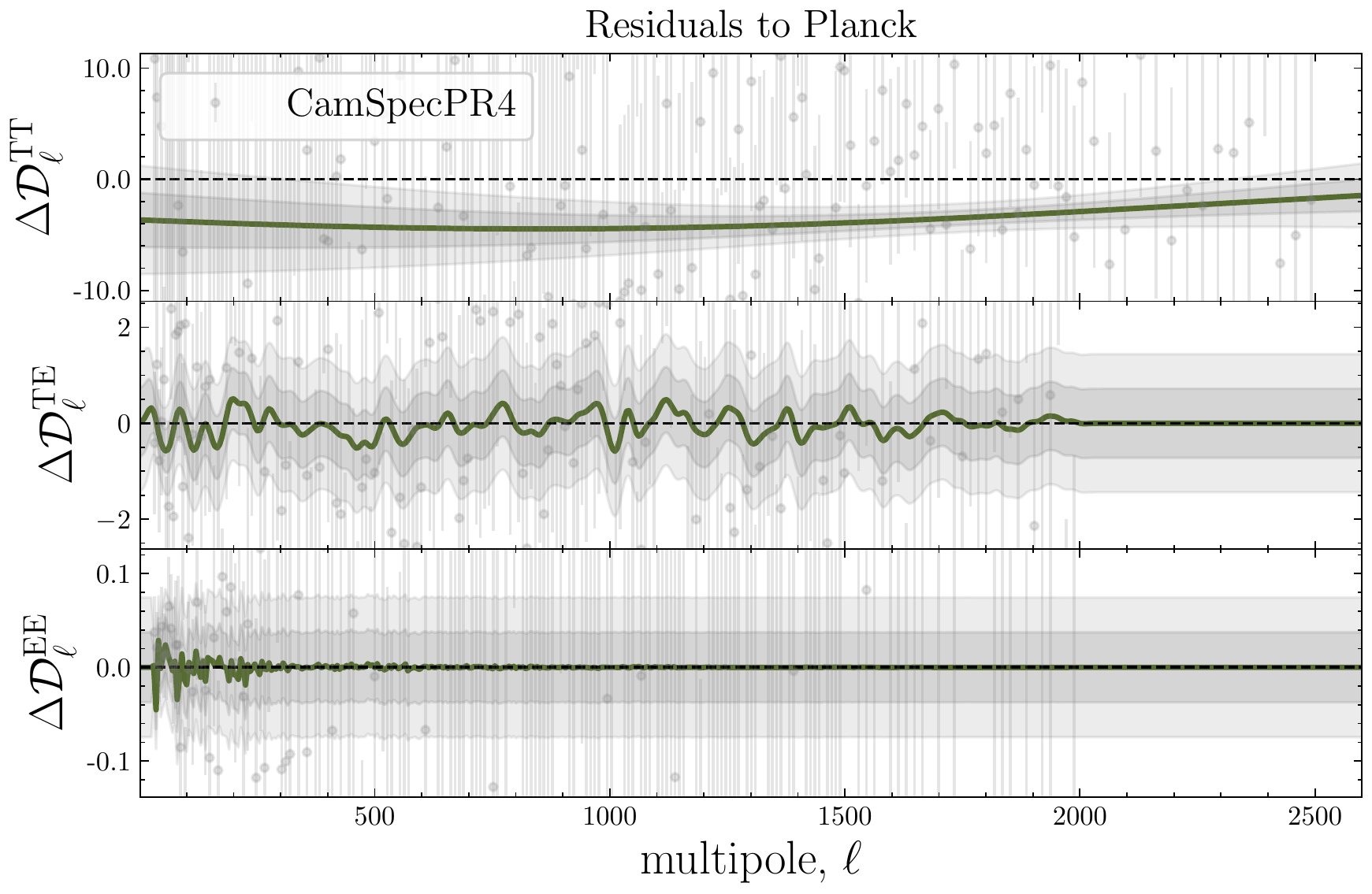}
    \includegraphics[width=0.475\columnwidth]{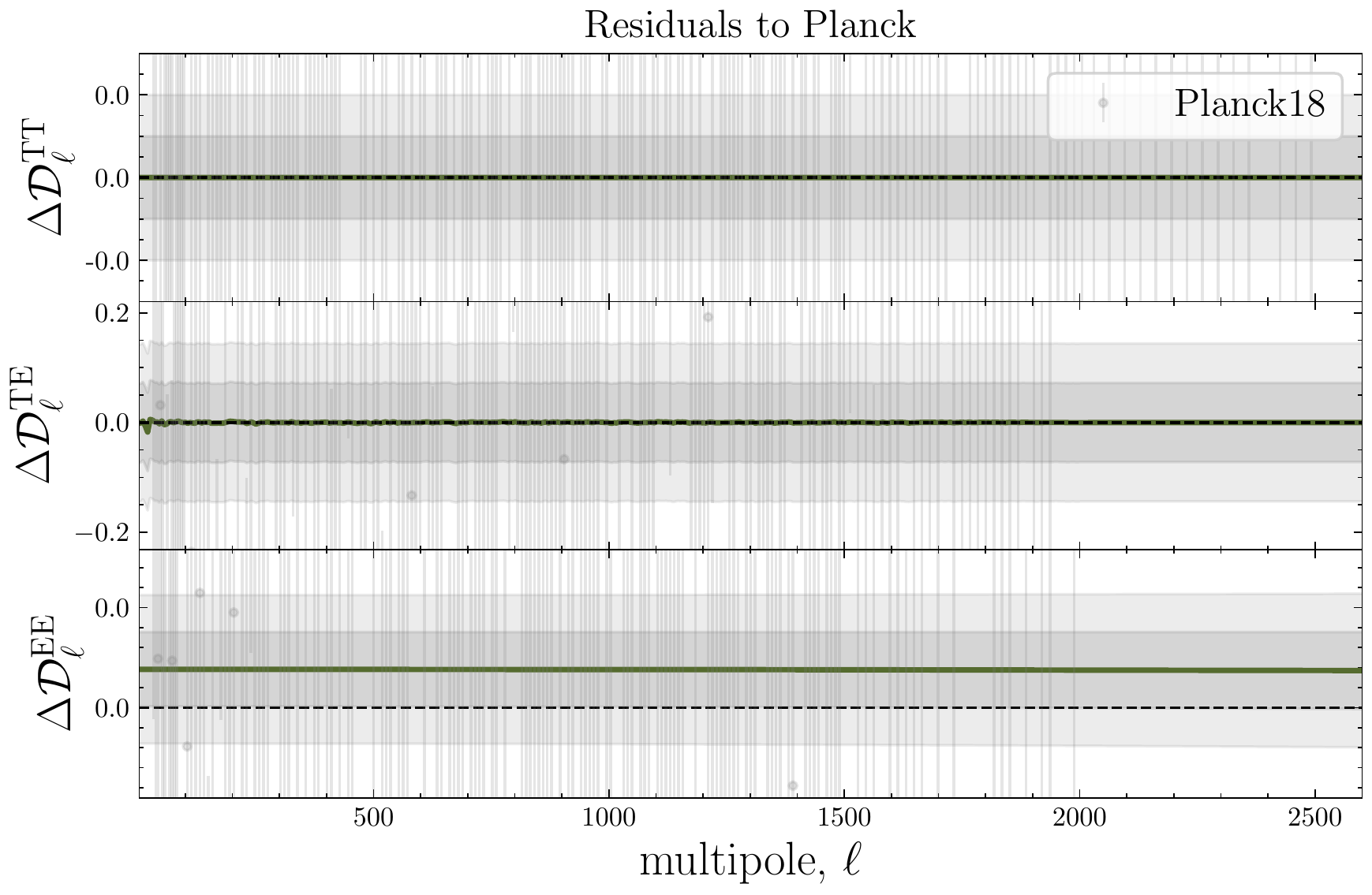}\\
    \includegraphics[width=0.475\columnwidth]{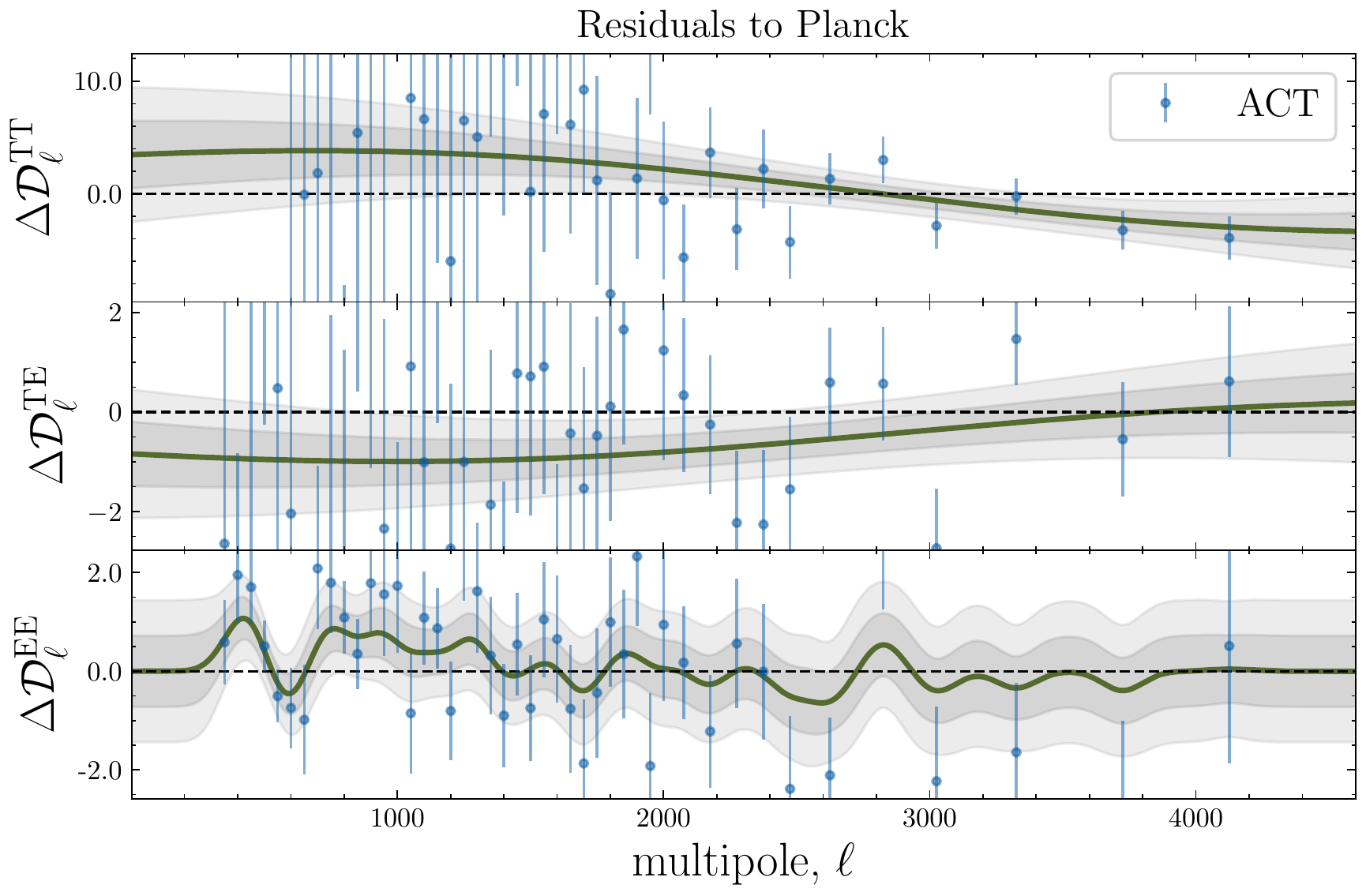}
    \includegraphics[width=0.475\columnwidth]{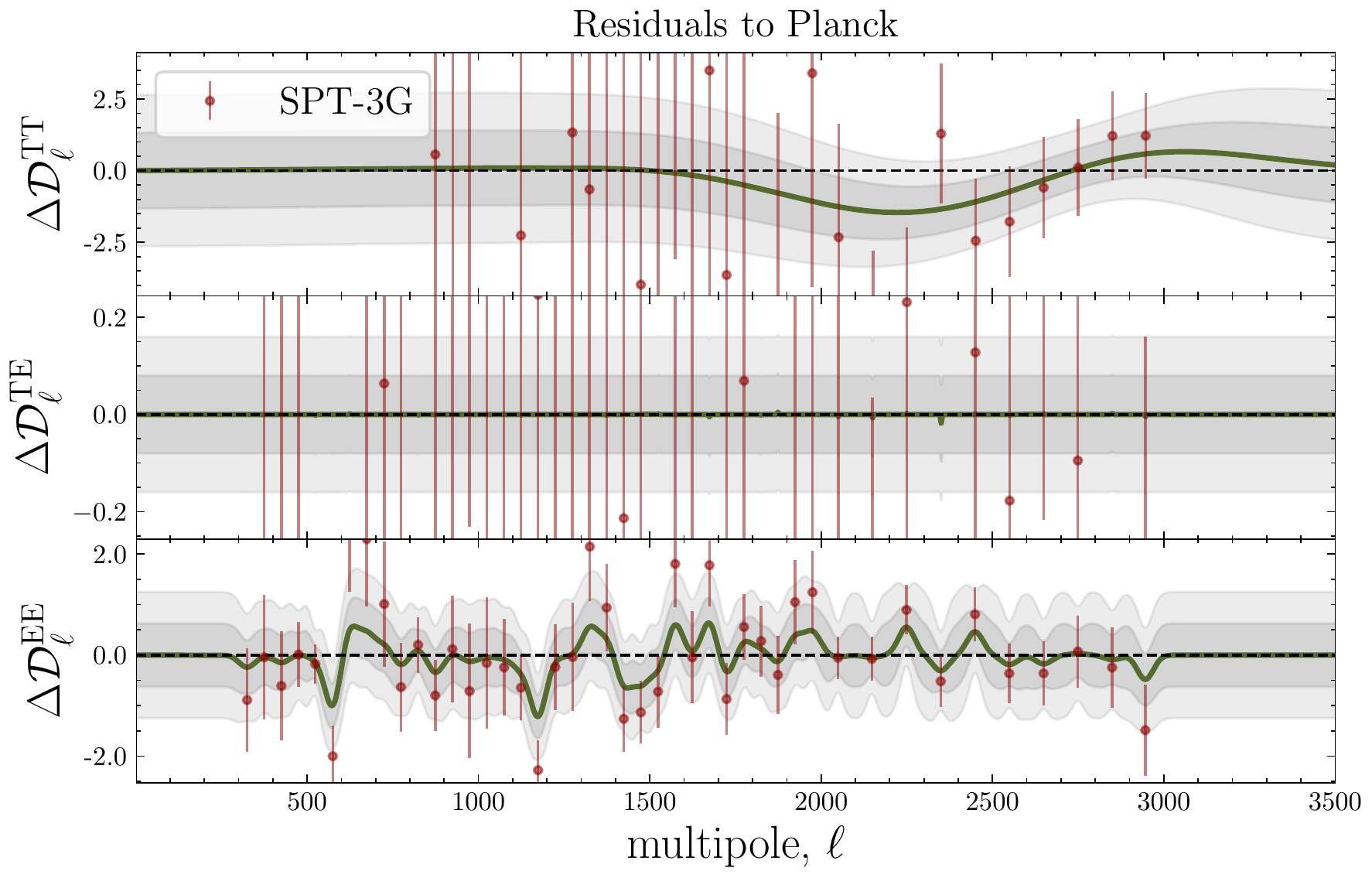}
    \caption{GP reconstructions of $\Delta\Dell\equiv\Dell^{\rm GP}-\Dell^{\Lambda\rm CDM}$ using the set of hyperparameters maximizing the log-marginal likelihood in Eq. \eqref{LML} when using the different datasets, specified in the legend. These results assume the \lcdm\ best-fit spectra to Planck18 (\texttt{Plik}) as the mean function. The solid line and shaded regions correspond to the mean and $2\sigma$ confidence intervals around it, respectively.}
    \label{fig:GP_residuals-P18}
\end{figure}

We now consider the \lcdm's best fit $\Dell$'s to the official Planck 2018 data release \cite{P18Cosmo2020} as the mean function in our analysis. The reconstructed residuals for the different datasets are shown in Fig. \ref{fig:GP_residuals-P18}. These are obtained using the set of hyperparameters \hyperpars\ maximizing the LML in Eq. \eqref{LML}, shown as black dots in Fig. \ref{fig:posteriors_P18-P18}. As noted before, the TT posteriors for CamSpec data (under the assumption of the best-fit \lcdm\ to \texttt{Plik}) reflect the differences in the temperature measurements from these two analyses, which might be non-parametric manifestations of the known $A_{L}$ and $\Omega_{k,0}$ anomalies \cite{Planck:2015fie,P18Cosmo2020,Schwarz:2015cma} which are no longer present in CamSpec \cite{Efstathiou_2021}; see also the discussion in Section \ref{P18-CamSpec}. The trend for the ACT reconstructed residuals $\Delta\Dell^{\rm TT}$ to favour large correlation lengths and less power at large scales (mimicking $n_s\to1$), observed in Section \ref{ACTDR4} remains the same, even when using the best fit to \texttt{Plik} as mean function. However, as shown in Fig \ref{fig:GP_residuals-P18}, the reconstructed TT residuals show milder (within $2\sigma$) deviations from zero, suggesting a slightly better agreement between ACT DR4 and \texttt{Plik} than with CamSpec. Although the TT measurements from SPT seem to be consistent with both CamSpec and \texttt{Plik} (see the TT panel in Fig. \ref{fig:heatmap}), the reconstructed residuals remain consistent with zero in the lower-right panel of Fig. \ref{fig:GP_residuals-P18}. An interesting aspect of this analysis is that of the reconstructed features in EE for both ACT and SPT, which seem to be stable under changes in the mean function. In other words, the EE posterior distributions for ACT and SPT are very similar, across Figs. \ref{fig:posteriors_ACT-CamSpec}, \ref{fig:Posteriors_SPT-CS}, \ref{fig:posteriors_P18-P18} and \ref{fig:posteriors_ACT-ACT+WMAP}, yielding similar reconstructed $\Delta\Dell$'s in Figs. \ref{fig:GP_ACT}, \ref{fig:GP_SPT},   \ref{fig:GP_residuals-P18} and \ref{fig:GP_residuals-ACT+WMAP}.

\begin{table}[h]
    \centering
    \caption{$\Delta\chisq=-2(\ln\Lkl^{\rm GP}-\ln\Lkl^{\Lambda \rm CDM})$ improvements in fit for the ground and space-based experiments obtained with the GP using \lcdm's best-fit to Planck 2018 as mean function.}
    {\rowcolors{2}{lgray}{white}
    \lcdm\ (\texttt{Plik}-Planck 2018)\\
    \begin{tabular}{cccccc}
        \hline
		$\Delta\chisq$ & \planck\ 2018 & CamSpec PR4  & ACT DR4 & SPT-3G \\ 
		\hline
		TT & $0$ & $-36.08$ & $-3.79$ & $-0.65$    \\ 
		TE & $0$ &$-0.93$ & $ -3.08$ & $-0.005$  \\ 
		EE & $-0.30$ & $-10.22$& $-6.94$ & $-9.30$  \\ 
		\hline
    \end{tabular}}
    \centering

    \label{tab:deltachi2_P18}
\end{table}

Finally, we would like to mention that the \planck\ analysis has been reproduced to $0.1\sigma$ accuracy using the new pipeline for the Simons Observatory \cite{SimonsPipeline}, providing an independent cross-check of the \planck\ results from \texttt{Plik}.

\subsection{Best-fit to \actpwmap}\label{sec:ACTWMAPmean}

\begin{figure}[t]
    \centering
    \includegraphics[width=0.45\columnwidth]{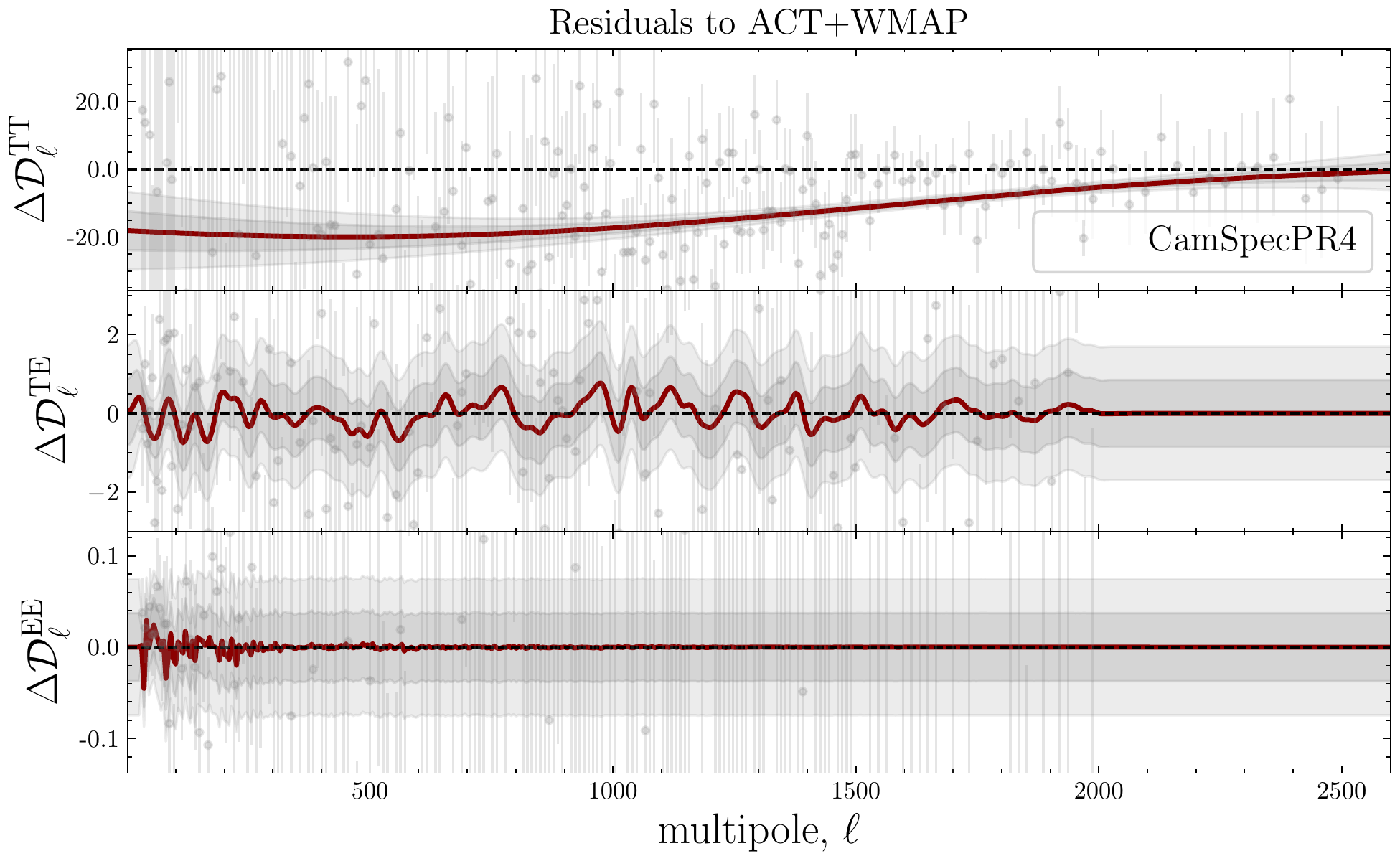}
    \includegraphics[width=0.45\columnwidth]{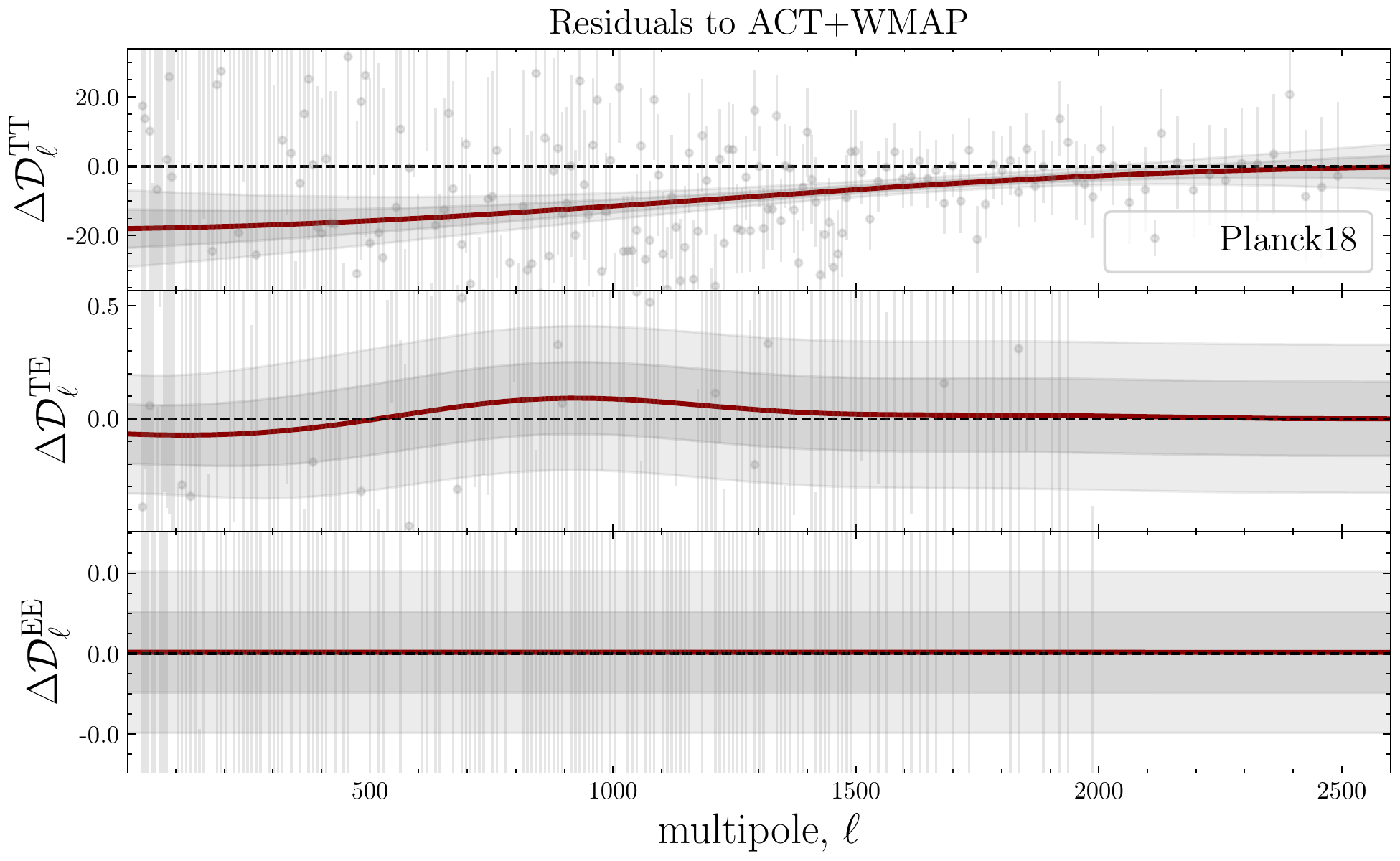}\\
    \includegraphics[width=0.45\columnwidth]{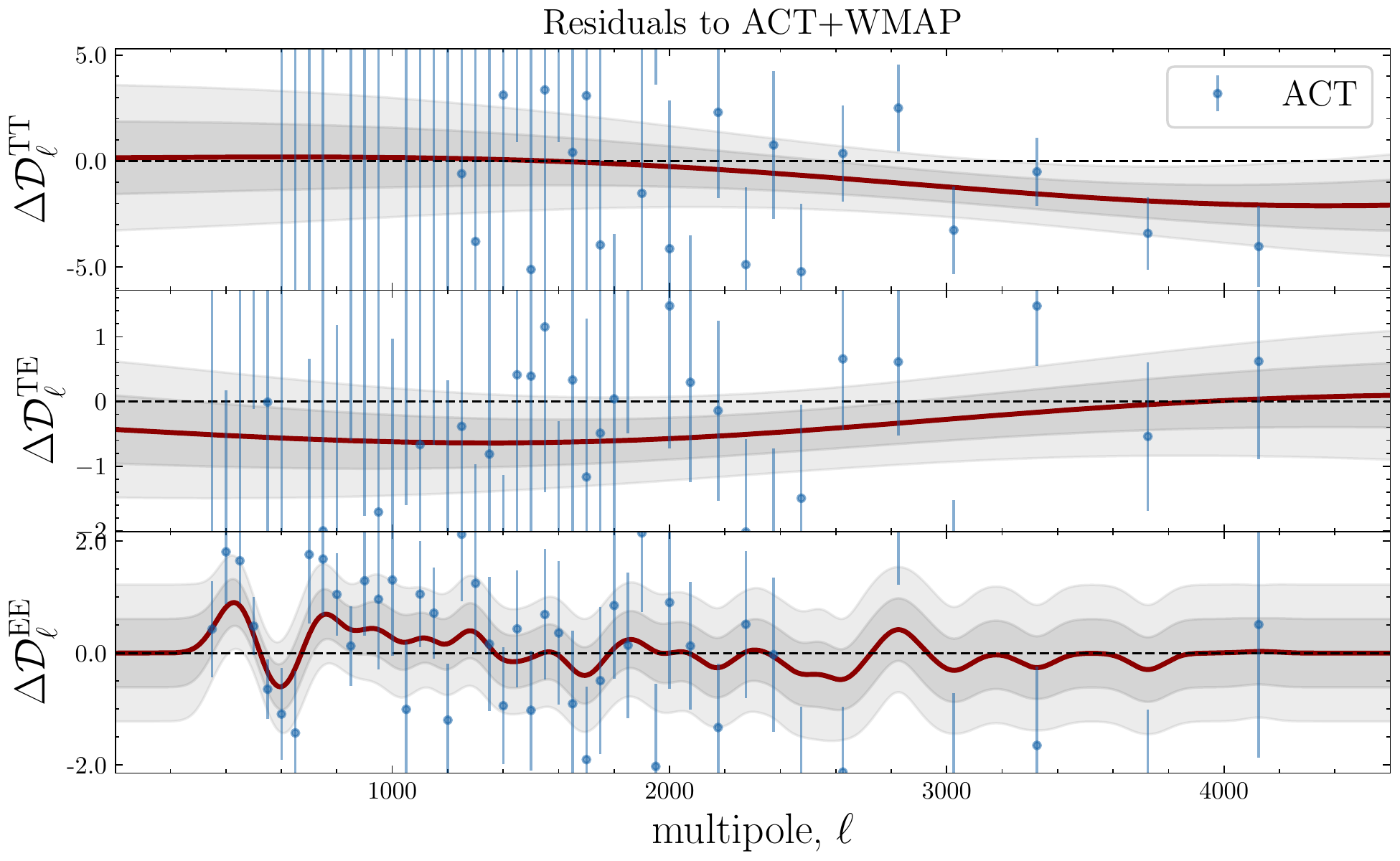}
    \includegraphics[width=0.45\columnwidth]{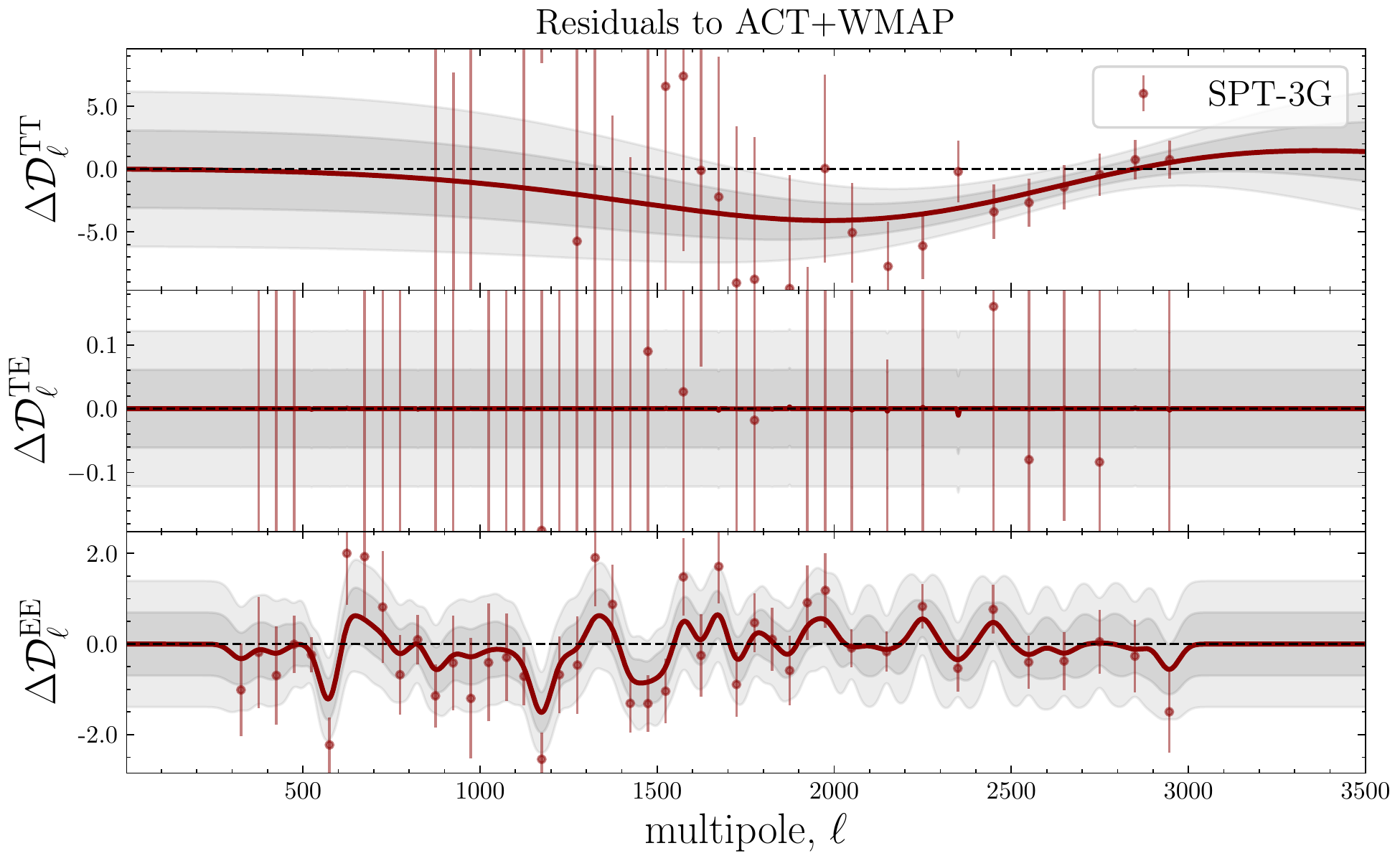}
    \caption{GP reconstructions of $\Delta\Dell\equiv\Dell^{\rm GP}-\Dell^{\Lambda\rm CDM}$ using the set of hyperparameters maximizing the log-marginal likelihood in Eq. \eqref{LML} when using the different datasets, specified in the legend. These results assume the \lcdm\ best-fit spectra to \actpwmap\ as the mean function. The solid line and shaded regions correspond to the mean and $2\sigma$ confidence intervals around it, respectively.}
    \label{fig:GP_residuals-ACT+WMAP}
\end{figure}

In Fig. \ref{fig:posteriors_ACT-ACT+WMAP}, we show the posteriors for \hyperpars\ when using yet another (different) mean function, namely \lcdm's best-fit to \actpwmap\ data\footnote{\url{ https://github.com/ACTCollaboration/pyactlike/blob/master/pyactlike/data/bf_ACTPol_WMAP_lcdm.minimum.theory_cl}} as given by Table 4 (fourth column) in \cite{Aiola_2020}. The top row shows the posteriors of the CamSpec and \texttt{Plik} residuals with respect to \actpwmap\ best-fit. The likelihood profiles for TE and EE are similar to the ones obtained using the \lcdm\ best fit to \planck; see Fig. \ref{fig:posteriors_CamSpec-CamSpec}. However, the posteriors for TT show a preference for even larger deviations from the mean function $\sigma_f\simeq10$, with significant improvements in fit found by the GP. This reflects the discordance between \planck\ TT measurements (both from \texttt{Plik} and CamSpec) and the \lcdm\ best fit to \actpwmap, as shown in Figs. \ref{fig:GP_residuals-ACT+WMAP} and \ref{fig:posteriors_ACT-ACT+WMAP}. More specifically, the TT reconstructions in Fig. \ref{fig:GP_residuals-ACT+WMAP} for both \texttt{Plik} and CamSpec show large ($\gg 3\sigma$) deviations from zero (dashed line), while the TE and EE measurements are consistent with the best-fit predictions from \actpwmap\ across the entire multipole range.\\

\begin{table}[h]
    \centering
    \caption{$\Delta\chisq=-2(\ln\Lkl^{\rm GP}-\ln\Lkl^{\Lambda \rm CDM})$ improvements in fit for the various datasets obtained with the GP using \lcdm's best-fit to ACT+WMAP as mean function.}
    \label{tab:deltachi2_ACT+WMAP}  
    \lcdm\ (ACT+WMAP)\\
    {\rowcolors{1}{white}{lgray}
    \begin{tabular}{ccccc}
        \hline
        \hline
		$\Delta\chisq$ &  \planck\ 2018 & CamSpecPR4  & ACT DR4 & SPT-3G \\ 
		TT & $-40.75$  &$-225.64$ & $-2.69$ &  $-7.469$ \\ 
		TE  & $-0.025$ & $ -1.15$ & $-1.31$ & $-0.0001$  \\ 
		EE  & $0  $ & $-10.20$ & $-4.28$ & $-13.735$ \\ 
		\hline
		\hline
    \end{tabular}}\\
    \centering
\end{table}
In the case of ACT DR4 data (third row in Fig. \ref{fig:posteriors_ACT-ACT+WMAP}), despite the large correlation lengths ($\ell_f\simeq10^3$ for TT \& TE and $\ell_f\simeq10^2$ for EE), the deviations from the mean function are mild $\sigma_f\simeq1$ for all cases. Curiously, the posteriors suggest (again) a mild bimodal distribution for \hyperpars\ in TT, TE, and EE with one mode located at $\hyperpars\simeq(1,10^3)$ and the second one at $(1,10^2)$. This seems to be the case, irrespective of the chosen mean function (best-fit to \planck\ or \actpwmap), and might suggest new features in the data that cannot be properly taken into account by a \lcdm\ model; see also the discussion on this in Section \ref{ACTDR4}. For comparison, we show the GP reconstructions of the ACT residuals with respect to \actpwmap\ best-fit predictions in Fig. \ref{fig:GP_residuals-ACT+WMAP}, to be compared with Fig. \ref{fig:GP_ACT} and \ref{fig:GP_residuals-P18} where the best-fit to CamSpec and \texttt{Plik} were used, respectively. The EE reconstructions show similar features irrespective of the chosen mean function (best-fit to Camspec, \planck\ or \actpwmap); see also the discussion in Section \ref{ACTDR4}. 
\\

Finally, we find that the posteriors for SPT-3G data in TT are consistent with the posteriors of the \planck\ data, (both from \texttt{Plik} and CamSpec) suggesting that the difference in TT is mainly coming from ACT. However, both ACT and SPT seem to agree on the bimodal distributions captured in EE. As discussed before, these particular values for \hyperpars\ lead to oscillations (features) around the \lcdm\ best-fit predictions, and ultimately to an improvement in fit with respect to \lcdm.
The improvements in fit in terms of $\Delta\chi^2$ values with respect to the mean-function (\lcdm\ best-fit to \actpwmap) are reported in Table \ref{tab:deltachi2_ACT+WMAP}.

\section{Supplementary Material:}
\subsection{\planck: High vs low-\tpdf{$\ell$}}\label{sec:HvsL}

\begin{figure}[h]
    \centering
    \includegraphics[width=0.32\textwidth]{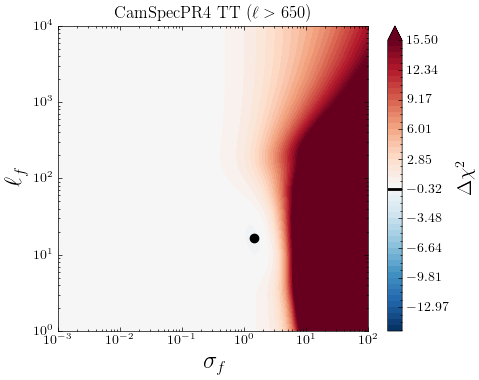}
    \includegraphics[width=0.32\textwidth]{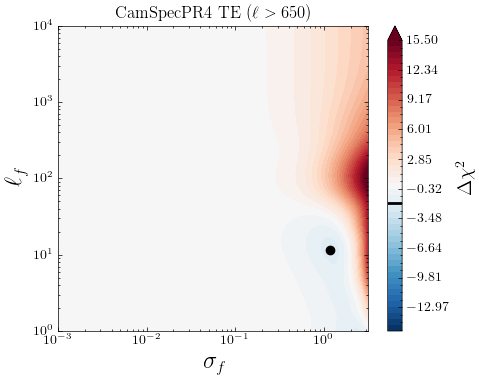}
    \includegraphics[width=0.32\textwidth]{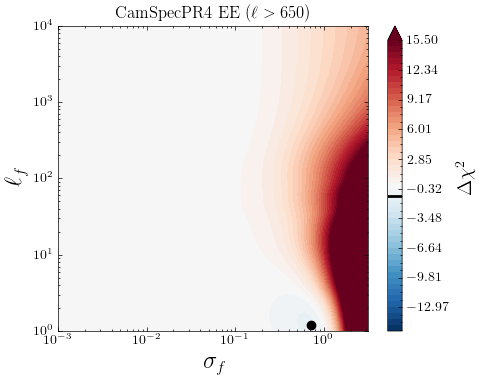}\\    \includegraphics[width=0.32\textwidth]{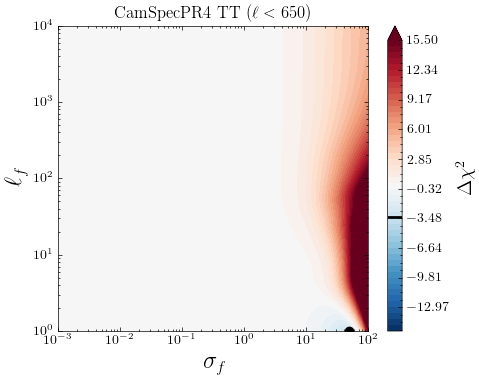}
    \includegraphics[width=0.32\textwidth]{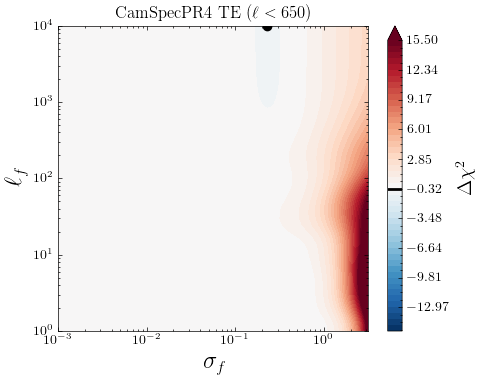}
    \includegraphics[width=0.32\textwidth]{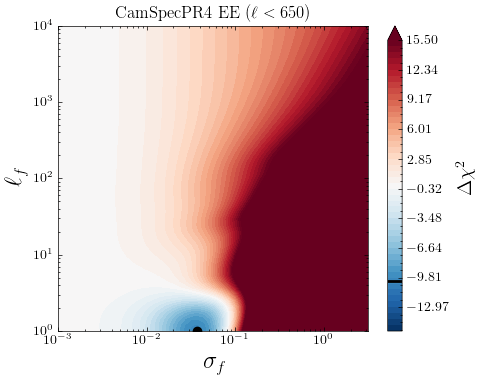}\\
    \includegraphics[width=0.32\textwidth]{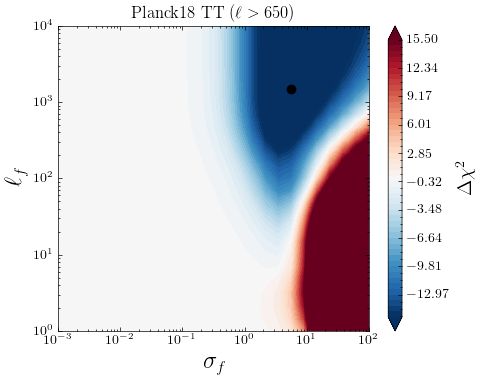}
    \includegraphics[width=0.32\textwidth]{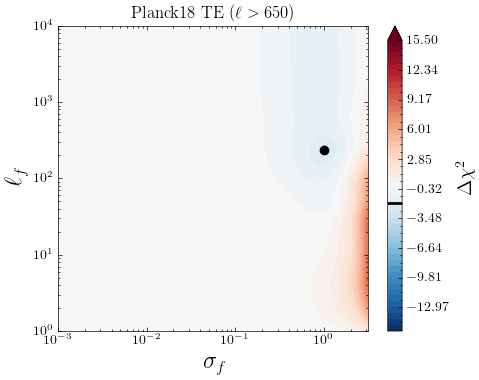}
    \includegraphics[width=0.32\textwidth]{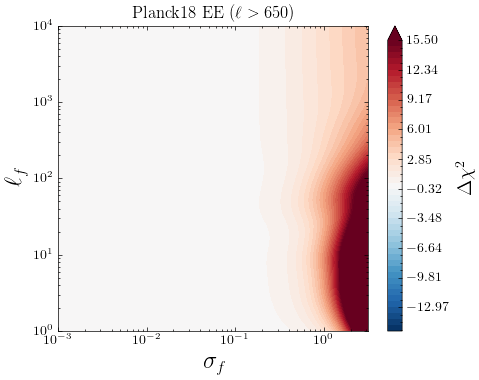}\\    \includegraphics[width=0.32\textwidth]{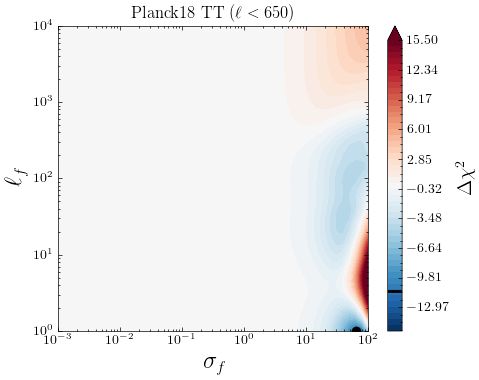}
    \includegraphics[width=0.32\textwidth]{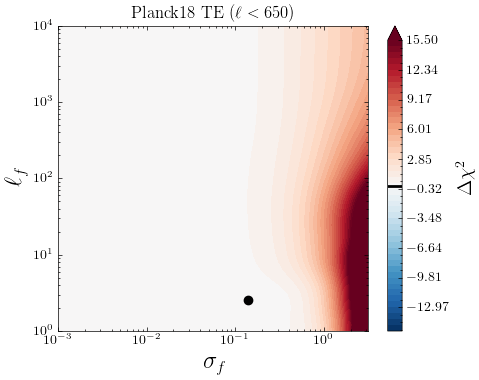}
    \includegraphics[width=0.32\textwidth]{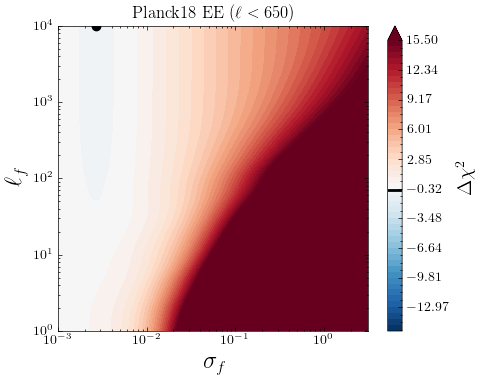}\\
    \caption{2D posterior distributions  as a function of \hyperpars\ for the various datasets, using \lcdm\ best-fit to CamSpec data as mean function.}
    \label{fig:posteriors_hvsl_CSPR4}
\end{figure}

Motivated by previous works \citep{Addison_2016,ACT_EDE,Poulin:2021bjr}, we study a low-high split of multipoles in the \planck\ data and see if there are indications of systematic errors specific to either of these parts. Fig \ref{fig:posteriors_hvsl_CSPR4} show the individual contributions from different $\ell$ ranges from the log marginal likelihoods for the CamSpec and \texttt{Plik} likelihoods, depicted in Figs \ref{fig:posteriors_CamSpec-CamSpec} and \ref{fig:posteriors_CamSpec-P18}, respectively.
We see that for the CamSpec best-fit mean function, the corresponding $\Delta\chi^2$ is small for TT measurements, and we find no significant dependence on the high/low-$\ell$ either. Little tendencies to prefer $\ell_f\lesssim1$ for low-$\ell$ is due to our Gaussian noise approximation slightly underestimating the non-Gaussian noise in low $\ell$s, for which the GP compensates by adding some uncorrelated noise. We conclude that no smooth deformation of the best-fit prediction is preferred to explain the original CamSpec data, and no low/high-$\ell$-specific systematics are detected in TT.
For the EE spectrum, we find that most of the improvements to the LML come from low-$\ell$ and around $\ell_f\lesssim 1$. This is consistent with our discussion in Section \ref{LCDM_CSPR4}; the CamSpec PR4 might have slightly underestimated the low-$\ell$ error, and the inclusion of uncorrelated random noise through GP ($\ell_f \lesssim 1$) noticeably improves the LML. Finally, as seen in Fig. \ref{fig:posteriors_hvsl_CSPR4}, the disagreement between \texttt{Plik} and CamSpec is mainly driven by the TT ($\ell>650$) part of the spectrum. The improvement in fit seen at smaller multipoles is likely due to the non-gaussian nature of the errors.

\subsection{Possible Calibration Issue}\label{Calibration}

\begin{figure}[h]
    \centering
    \includegraphics[width=0.32\textwidth]{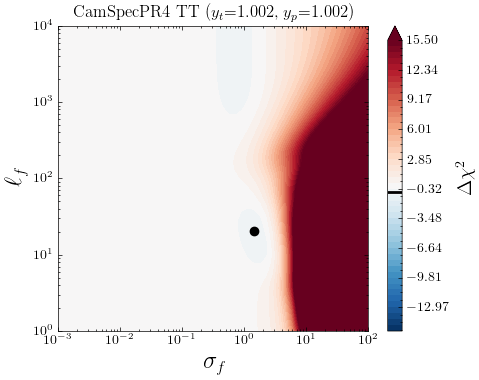}
    \includegraphics[width=0.32\textwidth]{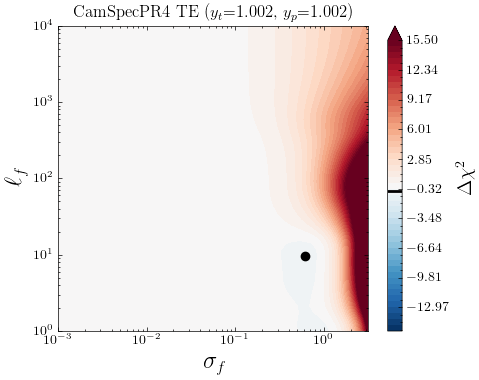}
    \includegraphics[width=0.32\textwidth]{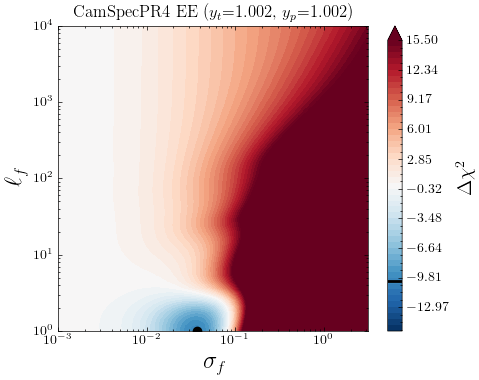}\\    \includegraphics[width=0.32\textwidth]{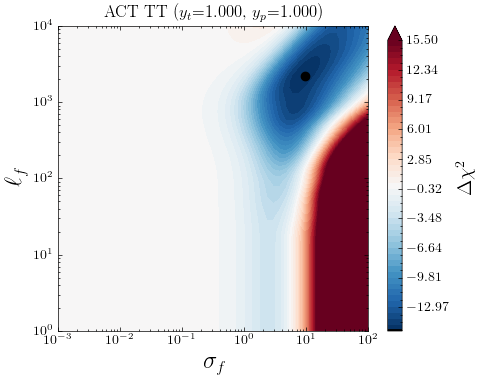}
    \includegraphics[width=0.32\textwidth]{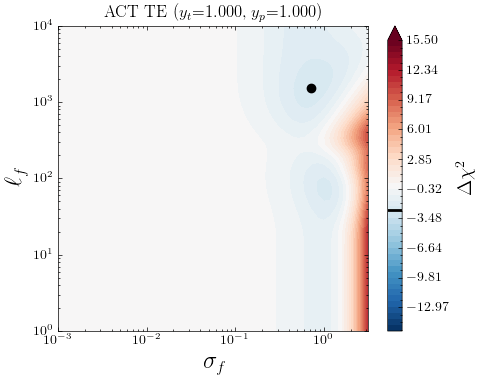}
    \includegraphics[width=0.32\textwidth]{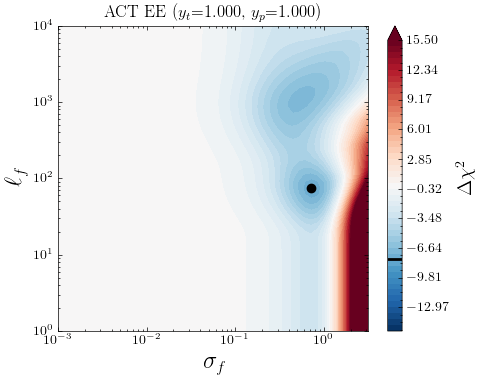}\\
    \caption{2D posterior distributions  as a function of \hyperpars\ for the various datasets, using the (scaled) \lcdm\ best-fit to CamSpec data as mean function. The value of the scaling parameters $(y_t, y_p)$ is reported in parenthesis on the top of each figure. The color bar shows the improvement in fit, where $\Delta\chisq=-2(\ln\Lkl^{\rm GP}-\ln\Lkl^{\Lambda \rm CDM})$.}
    \label{fig:posteriors_Scaled-CamSpec}
\end{figure}

\begin{figure}[p]
    \centering
    \includegraphics[scale=0.45]{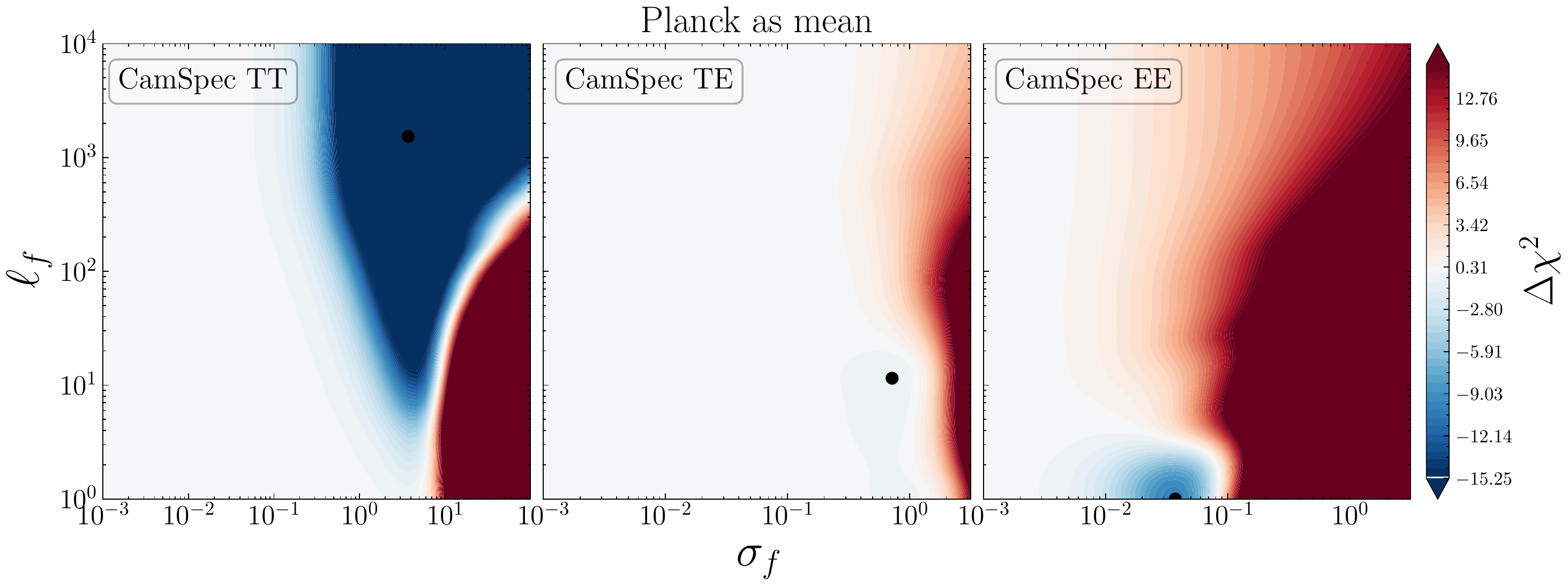}
    \includegraphics[scale=0.45]{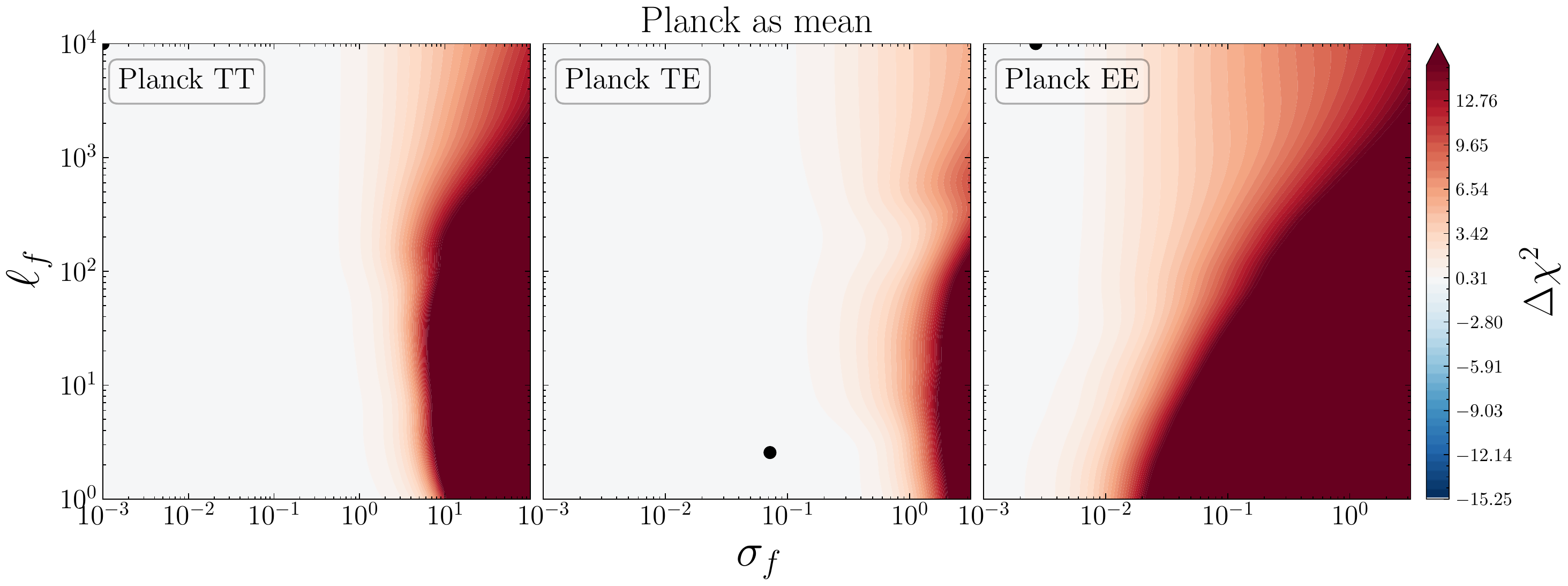}
    \includegraphics[scale=0.45]{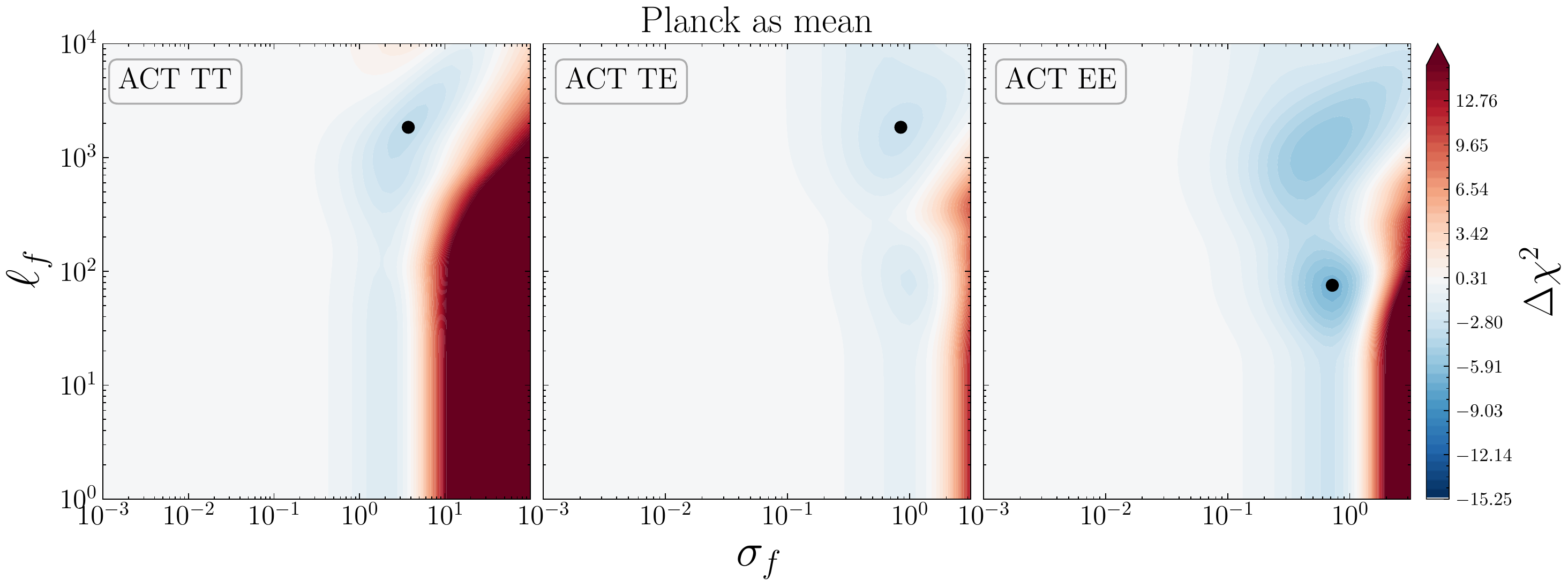}
    \includegraphics[scale=0.45]{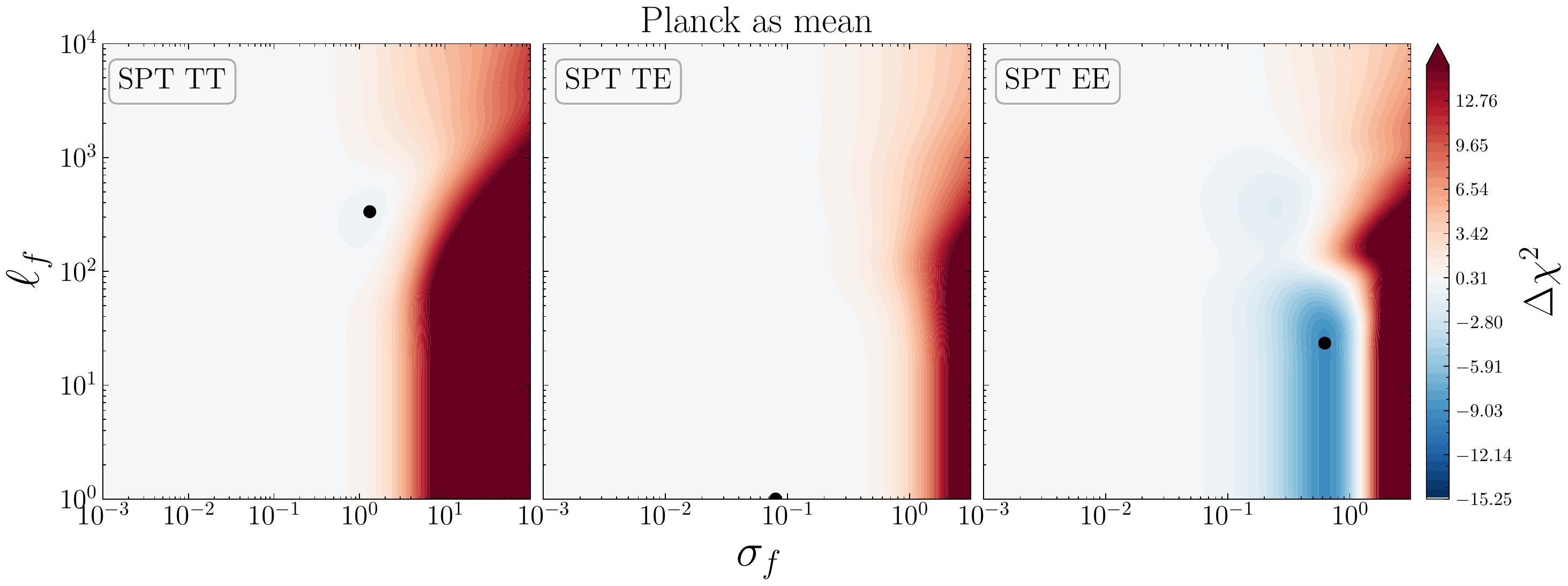}
    \caption{2D posterior distributions as a function of \hyperpars\ for the various datasets, using the \lcdm\ best-fit to the \planck (\texttt{Plik}) data as mean function. The color bar shows the improvement in fit, where $\Delta\chisq=-2(\ln\Lkl^{\rm GP}-\ln\Lkl^{\Lambda \rm CDM})$ and $\ln\Lkl^{\rm GP}\hyperpars$ is the log marginal likelihood in Eq. \eqref{LML}.}
    \label{fig:posteriors_P18-P18}
\end{figure}

\begin{figure}[p]
    \centering
    \includegraphics[scale=0.45]{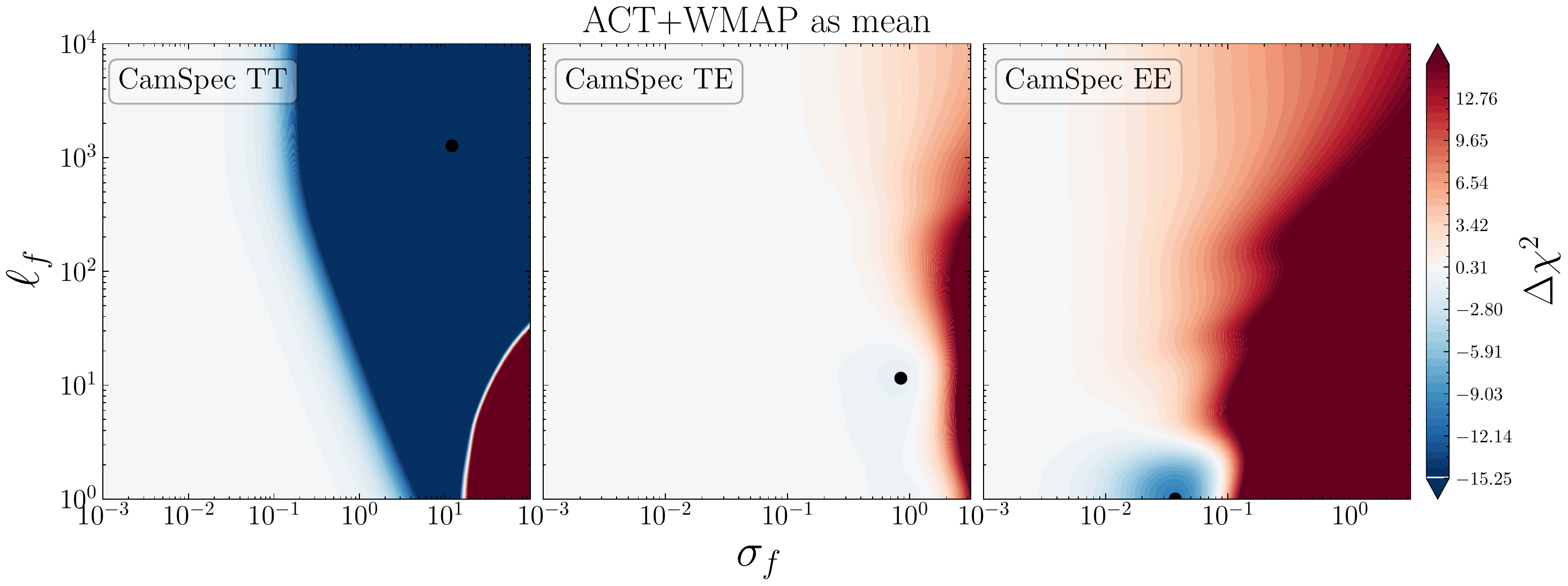}    
    \includegraphics[scale=0.45]{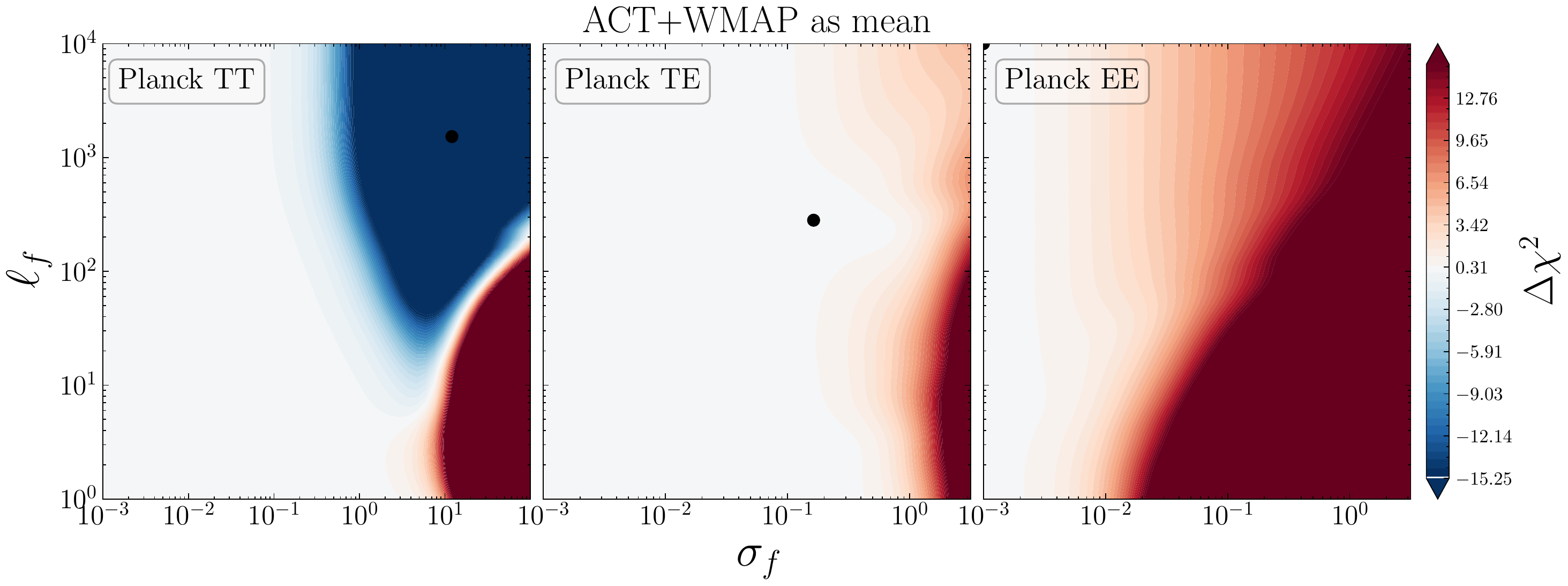}
    \includegraphics[scale=0.45]{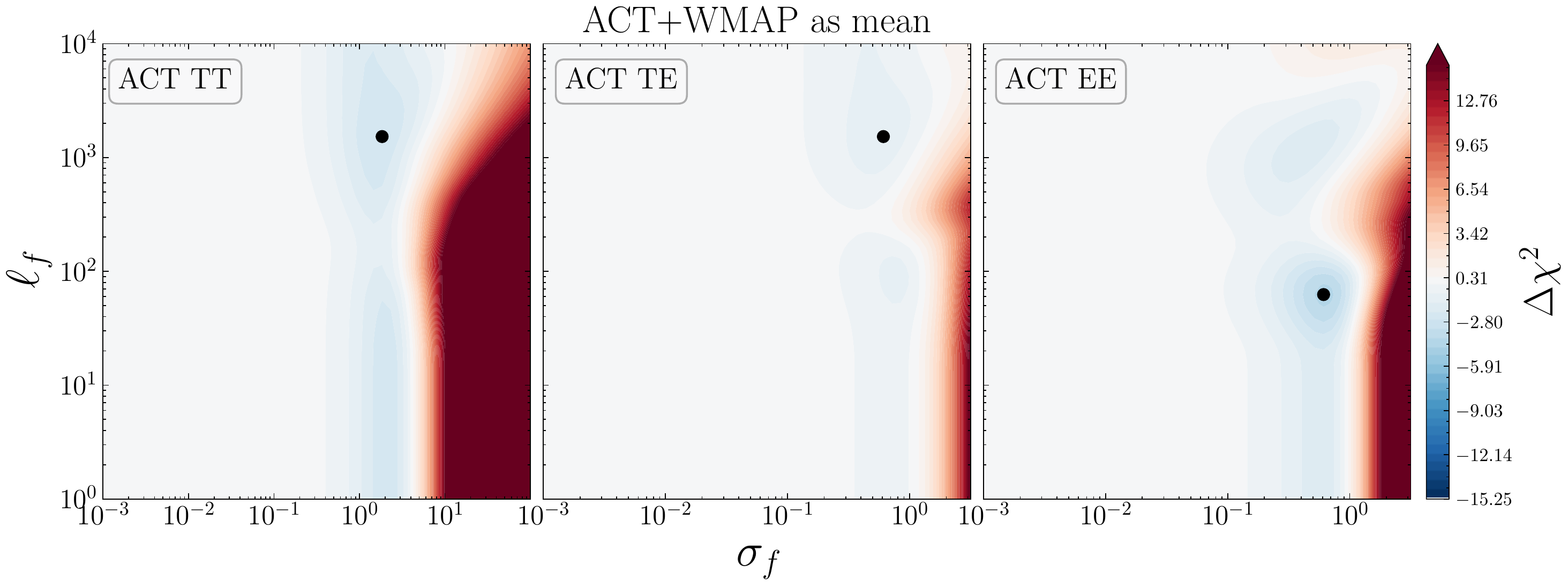}
    \includegraphics[scale=0.45]{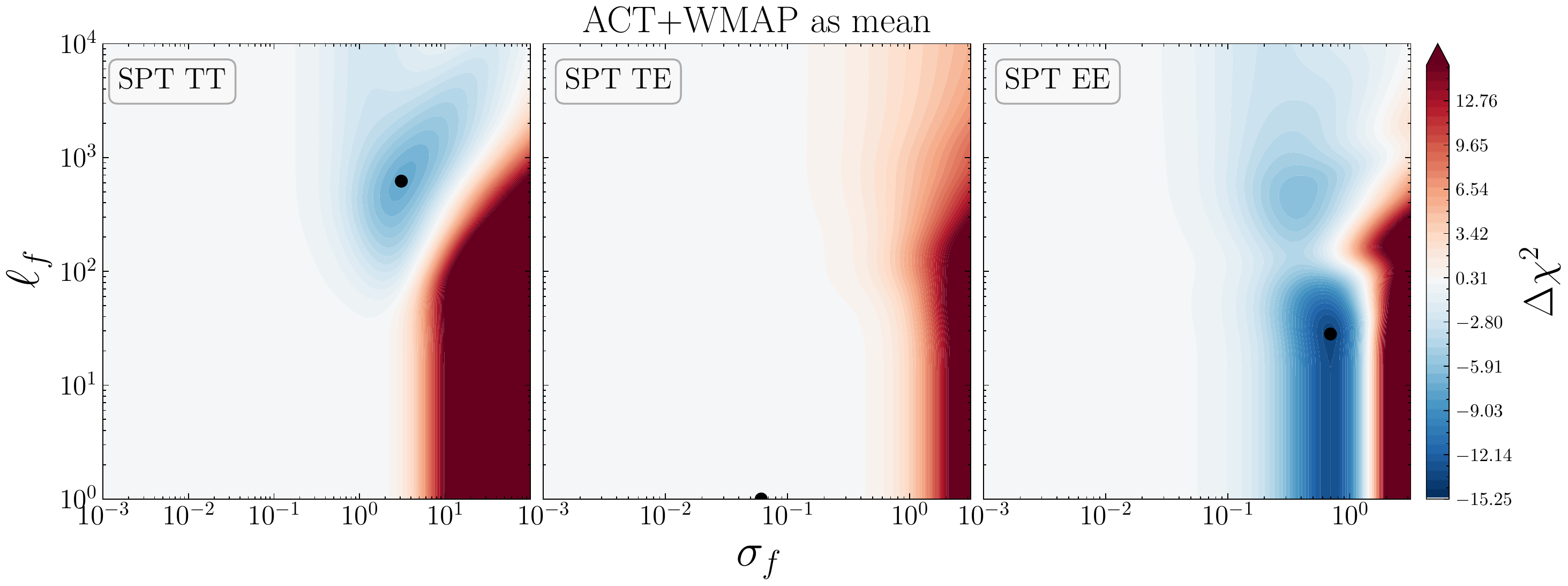}
    \caption{2D posterior distributions  as a function of \hyperpars\ for the various datasets, using \lcdm\ best-fit to \actpwmap\ data as mean function. The color bar shows the improvement in fit, where $\Delta\chisq=-2(\ln\Lkl^{\rm GP}-\ln\Lkl^{\Lambda \rm CDM})$.}
    \label{fig:posteriors_ACT-ACT+WMAP}
\end{figure}

To account for possible differences in the absolute magnitude calibration of the CMB maps, we introduce temperature and polarisation scaling parameters $y_t$ and $y_p$, which are close to unity. This is analogous to the nuisance parameters introduced by the \planck\ and ACT collaborations, in terms of $A_{\rm Planck}$ and $y_p$, respectively. These parameters are multiplying the theoretical spectrum such that:
\begin{align}
    \Dell^{\rm TT,obs} & = y_{t}^2\cdot\Dell^{\rm TT,th},\nonumber \\
    \Dell^{\rm TE,obs} & = y_t\cdot y_{p}\cdot\Dell^{\rm TE,th},\\
    \Dell^{\rm EE,obs} & = y_{p}^2\cdot\Dell^{\rm EE,th}\nonumber.
\end{align}
The values of these parameters are chosen such that the combined $\chi^2_{\rm tot}(y_t,y_p)=\chisq_{\rm TT}(y_t)+\chisq_{\rm TE}(y_t,y_p)+\chisq_{\rm EE}(y_p)$ is minimized. In the ideal case, one would expect to have $y_t=y_p=1$. As an example, let us note that the best-fit spectra provided by the \planck\ collaboration is to be scaled by an overall calibration parameter $A_{\rm Planck}\equiv1/\texttt{calPlanck}^2$ for the theoretical predictions to be properly compared with the coadded spectra. 
Thus, for a given \lcdm\ theoretical \Dell, we find the value of $y_{t,p}$ minimizing the \chisq\ to the observed data and proceed to subtract the \emph{(scaled)} theoretical spectrum from the \emph{actual observed} data. In practice, we use scipy's ``differential evolution'' (stochastic) minimizer \cite{2020SciPy-NMeth} to find the optimal values for $y_{t,p}$. We then apply our GP formalism and inspect for changes in the posteriors under such scaling of the spectra. The results are shown in Fig. \ref{fig:posteriors_Scaled-CamSpec}, where we report the values of the scaling parameters ($y_t$ and $y_p$, minimizing the combined 
\chisq) in parenthesis on top of each figure.
For other consistency checks and discussion on possible scaling issues in the data, we refer the reader to \citep[e.g.][]{Hazra_2014,Shafieloo_2017,Laposta22}.

\bibliographystyle{JHEP}
\bibliography{ref}
\end{document}